\documentclass[journal]{IEEEtran}
\usepackage{amsmath,amsfonts}
\usepackage{algorithmic}
\usepackage{array}
\usepackage[caption=false,font=normalsize,labelfont=sf,textfont=sf]{subfig}
\usepackage{textcomp}
\usepackage{stfloats}
\usepackage{url}
\usepackage{verbatim}
\usepackage{graphicx}
\usepackage{xcolor}
\usepackage[short,c2,nocomma]{optidef}
\usepackage{hyperref}
\newcommand{\mytitle}{Reinforcement Learning with Model Predictive Control for Highway Ramp Metering}
\hypersetup{
    hidelinks,
    pdfauthor=Filippo Airaldi,
    pdfcreator=Filippo Airaldi,
    pdftitle=\mytitle,
}
\usepackage{cleveref}
\crefformat{figure}{Fig.~#2#1#3}
\Crefformat{figure}{Fig.~#2#1#3}
\usepackage[inline]{enumitem}
\usepackage{tikz}
\usepackage[utf8]{inputenc}
\usepackage{pgfplots}
\DeclareUnicodeCharacter{2212}{−}
\usepgfplotslibrary{groupplots,dateplot}
\usetikzlibrary{patterns,shapes.arrows}
\pgfplotsset{compat=newest}
\def\axisdefaultheight{110pt}
\usepackage{multirow}
\usepackage{bm}
\usepackage{siunitx}
\DeclareSIUnit{\vehicle}{veh}
\DeclareSIUnit{\lane}{lane}

\DeclarePairedDelimiter{\norm}{\lVert}{\rVert}
\DeclarePairedDelimiter{\card}{\lvert}{\rvert}
\DeclarePairedDelimiter{\stochasticargument}{[}{]}
\NewDocumentCommand{\expval}{e{_}}{%
    \mathbb{E}\IfValueT{#1}{_{#1}}\stochasticargument
}
\providecommand\given{}
\DeclarePairedDelimiterXPP{\probab}[1]{\mathbb{P}}{[}{]}{}{
\renewcommand\given{\nonscript\:\delimsize\vert\nonscript\:\mathopen{}}
#1}
\newtheorem{remark}{Remark}
\def\arxiv{1}

\title{ \mytitle }
\author{
    Filippo Airaldi,
    Bart De Schutter, \ifx\arxiv\undefined\IEEEmembership{Fellow, IEEE}, \fi
    Azita Dabiri

    \thanks{
        \ifx\arxiv\undefined
            Manuscript received XX XX, XX; revised XX XX, XX; accepted XX XX, XX. Date of publication XX XX, XX; date of current version XX XX, XX.
        \fi
        This research is part of a project that has received funding from the European Research Council (ERC) under the European Union’s Horizon 2020 research and innovation programme (Grant agreement No. 101018826 - CLariNet).
        \ifx\arxiv\undefined
            The Associate Editor for this paper was XX.
        \fi
        (\textit{Corresponding author: Filippo Airaldi}.)
    }
    \thanks{
        The authors are with the Delft Center for Systems and Control, Delft University of Technology, Mekelweg 2, 2628 CD Delft, The Netherlands (e-mail: f.airaldi@tudelft.nl; b.deschutter@tudelft.nl; a.dabiri@tudelft.nl).
    }
    \ifx\arxiv\undefined
        \thanks{Digital Object Identifier XX}
    \fi
}

\ifx\arxiv\undefined
\fi

\begin{document}

\maketitle

\begin{abstract}
    In the backdrop of an increasingly pressing need for effective urban and highway transportation systems, this work explores the synergy between model-based and learning-based strategies to enhance traffic flow management by use of an innovative approach to the problem of ramp metering control that embeds Reinforcement Learning (RL) techniques within the Model Predictive Control (MPC) framework. The control problem is formulated as an RL task by crafting a suitable stage cost function that is representative of the traffic conditions, variability in the control action, and violations of the constraint on the maximum number of vehicles in queue. An MPC-based RL approach, which leverages the MPC optimal problem as a function approximation for the RL algorithm, is proposed to learn to efficiently control an on-ramp and satisfy its constraints despite uncertainties in the system model and variable demands. Simulations are performed on a benchmark small-scale highway network to compare the proposed methodology against other state-of-the-art control approaches. Results show that, starting from an MPC controller that has an imprecise model and is poorly tuned, the proposed methodology is able to effectively learn to improve the control policy such that congestion in the network is reduced and constraints are satisfied, yielding an improved performance that is superior to the other controllers. 
\end{abstract}

\begin{IEEEkeywords}
    Ramp Metering, Reinforcement Learning, Model Predictive Control.
\end{IEEEkeywords}

\section{Introduction} \label{sections:introduction}

\IEEEPARstart{O}{ver} the past decades, the need for urban and highway transportation has become significantly relevant to our society. Extended travel times (primarily spent in traffic jams), impact on stress levels and pollution are but a few examples of the negative effects due to high demands for mobility that traffic engineers are facing nowadays \cite{johansson_2014_blueprint,who_2018_global}. In particular, the necessity for advanced intelligent traffic control strategies is arising as the construction of new infrastructure and upgrades to the existing one become increasingly unsustainable, both in economical, environmental, and societal terms \cite{siri_2021_freeway,castaneda_2022_highway}.

Ramp metering (RM) control has been proven to be an effective tool in regulating the traffic flow entering the network at on-ramps \cite{papageorgiou_2002_freeway}. It consists in modulating the flow at an on-ramp via traffic lights with appropriate timings of red, green, and amber lights. Earlier approaches were fixed-time, i.e., the timings were calculated offline with the help of historic traffic data \cite{wattleworth_1965_peak}. More efficient strategies involve real-time measurements of the traffic conditions, enabling online computation of the timings in a traffic-responsive way. In these cases, the underlying algorithms range from simpler feedback strategies \cite{papageorgiou_1991_alinea,wang_2014_local,frejo_2019_feedforward} to more complex architectures, such as Model Predictive Control (MPC) \cite{hegyi_2002_optimal,hegyi_2004_model,hegyi_2005_model} and Reinforcement Learning (RL) \cite{wang_2022_integrated,han_2022_physics}. One of the primary research challenges in RM today is the design of strategies that can control the entering flow in a way that effectively balances congestion in the freeway and the queue length at the on-ramp, two often conflicting goals \cite{shaaban_2016_literature,vrbanic_2021_variable}. Most strategies deployed on actual ramps are either feedback-based local policies, which are inherently myopic, or model-based policies and, as such, are vulnerable to model inaccuracies and thus require precise identification of traffic parameters, as well as an attentive assessment of their efficacy \cite{ma_2020_statistical}. Conversely, data-driven approaches, which are one of the main focuses of current research, typically necessitate extensive datasets of real or synthetic traffic data for training.

In modern control literature, MPC is an established control paradigm whose appeal can be attributed to its strong theoretical foundations, a diverse array of applications, as well as its remarkable capability to manage multivariate and constrained systems \cite{rawlings_2017_model}. Its versatility further allows for alternative formulations that can systematically handle disturbances affecting the system, e.g., in a robust \cite{bemporad_1999_robust} or a stochastic \cite{mesbah_2016_stochastic} manner. Unlike other black-box or purely data-driven approaches, MPC-based control strategies often maintain a high degree of explainability and interpretability, which is especially favourable in settings where constraint satisfaction is critical. These properties are in part thanks to the prediction model that the MPC scheme contains, which allows it to forecast the dynamical evolution of state trajectories along the time horizon, and which is often designed by domain experts. 

Nonetheless, it is well known that the performance of such predictive controllers is heavily bound to the quality of the prediction model, which calls for a high level of expertise during the system identification phase. In an attempt to counter this limitation, driven by the recent surge in advancements in Machine Learning (ML) algorithms and facilitated by the growing computational and sensing capabilities of digital systems, there is currently a substantial body of research dedicated to learning-based control methodologies that offer ways of leveraging open- or closed-loop data to, e.g., synthesize models and controllers, or enhance existing ones.

Amongst these methodologies, one of the most notable approaches is Reinforcement Learning (RL), a paradigm of ML that has gained substantial attention due to its advancements and successes. It provides a framework that allows to train an agent to interact with an unknown environment and to make decisions to maximise rewards or minimise costs \cite{sutton_2018_reinforcement}. The popularity in the contemporary ML community tends to favour its model-free variants that rely solely on observed state transitions and stage cost realisations to increase the performance of the control policy. Early RL focused on tabular approaches that enjoy simplicity, some degree of interpretability, and convergence guarantees, but they are only applicable to small and simple environments \cite{watkins_1992_qlearning}, and are often limited in handling larger state spaces, exhibit slow convergence, and struggle to generalize. As the field progressed, researchers recognized the potential of function approximators and, in particular, of Neural Networks (NNs) \cite{busoniu_2017_reinforcement}, leading to the development of Deep RL (DRL). However, while DRL has demonstrated great efficacy \cite{mnih_2013_playing}, its NN-based policies generally lack interpretability, and the integration of prior knowledge (e.g., from domain experts) is still difficult. Thus, providing formal guarantees on properties relevant to a controller, such as stability and constraint satisfaction, in the context of NN-enhanced control architectures is one of today's challenges in the field. Additionally, existing model-free approaches are usually plagued by sample inefficiency and poor real-world performance, since they are often trained offline in approximate simulators and require large amounts of data for meaningful converge \cite{dulac_2021_challenges}.

At the same time, on the other side of the coin, the control community is also dedicating efforts in integrating data-driven techniques with control systems in more structured ways, fostering methodologies that aim to learn various components of a controller directly from data \cite{hewing_2020_learning,brunke_2022_safe}. The goals include, e.g., reducing model uncertainties and the impact of disturbances, improving closed-loop performance and battling conservativeness, or promoting robust control policies with guarantees on constraint satisfaction. Notably, learning-based MPC methodologies \cite{mesbah_2022_fusion} have been devised to, e.g., leverage Gaussian Process regression to learn unmodelled nonlinear dynamics in the prediction model via state transition observations \cite{hewing_2020_cautious}, or use Bayesian Optimisation to automatically tune and optimise the controller hyperparameters \cite{piga_2019_performance}. In the literature, MPC has been successfully combined also with RL, e.g., as a safety filter for the learning-based agents (also called \textit{shielding}) \cite{hewing_2020_learning,wabersich_2022_probabilistic}, as a model reference and baseline control provider \cite{romero_2024_actorcritic,zhang_2022_modelreference}, and to solve mixed-logical control problems more efficiently \cite{dasilva_2024_integrating}.

\begin{figure}
    \centering
    \begin{tikzpicture}[x=0.75pt,y=0.75pt,yscale=-0.8175,xscale=1]
    \draw (140,89) -- (140,79) -- (212.93,2.18);
    \draw [shift={(215,0)}, rotate = 133.51, fill={rgb,255:red,0;green,0;blue,0}, line width=0.08, draw opacity=0] (10.72,-5.15) -- (0,0) -- (10.72,5.15) -- (7.12,0) -- cycle;
    \draw (109,160) -- (70,160) -- (70,110) -- (91,110);
    \draw [shift={(94,110)}, rotate = 180, fill={rgb,255:red,0;green,0;blue,0}, line width=0.08, draw opacity=0] (10.72,-5.15) -- (0,0) -- (10.72,5.15) -- (7.12,0) -- cycle;
    \draw (109,170) -- (51,170) -- (51,40) -- (96,40) ;
    \draw [shift={(99,40)}, rotate=180, fill={rgb,255:red,0;green,0;blue,0}, line width=0.08, draw opacity=0] (10.72,-5.15) -- (0,0) -- (10.72,5.15) -- (7.12,0) -- cycle;
    \draw (240,40) -- (290,40) -- (290,165) -- (232,165) ;
    \draw [shift={(229,165)}, rotate=360, fill={rgb,255:red,0; green,0;blue,0}, line width=0.08, draw opacity=0] (10.72,-5.15) -- (0,0) -- (10.72,5.15) -- (7.12,0) -- cycle;

    \draw  [color={rgb,255:red,254;green,151;blue,43}, draw opacity=1, fill={rgb,255:red,255;green,230;blue,204}, fill opacity=1] (95,98) .. controls (95,93.58) and (98.58,90) .. (103,90) -- (177,90) .. controls (181.42,90) and (185,93.58) .. (185,98) -- (185,122) .. controls (185,126.42) and (181.42,130) .. (177,130) -- (103,130) .. controls (98.58,130) and (95,126.42) .. (95,122) -- cycle ;
    \draw [color={rgb,255:red,93;green,195;blue,79}, draw opacity=1, fill={rgb,255:red,213;green,232;blue,212}, fill opacity=1] (109,153) .. controls (109,148.58) and (112.58,145) .. (117,145) -- (221,145) .. controls (225.42,145) and (229,148.58) .. (229,153) -- (229,177) .. controls (229,181.42) and (225.42,185) .. (221,185) -- (117,185) .. controls (112.58,185) and (109,181.42) .. (109,177) -- cycle;
    \draw [color={rgb,255:red,87;green,113;blue,214}, draw opacity=1, fill={rgb,255:red,218;green,232;blue,252}, fill opacity=1] (100,28) .. controls (100,23.58) and (103.58,20) .. (108,20) -- (232,20) .. controls (236.42,20) and (240,23.58) .. (240,28) -- (240,52) .. controls (240,56.42) and (236.42,60) .. (232,60) -- (108,60) .. controls (103.58,60) and (100,56.42) .. (100,52) -- cycle;

    \draw (140,110) node [align=left] {\textbf{RL Agent}};
    \draw (169,165) node [align=left] {\textbf{Environment}};
    \draw (170,40) node [align=left] {\textbf{Parametric MPC}};
    \draw (54,82) node [anchor=north west, inner sep=0.75pt, align=left] {state};
    \draw (152,67) node [anchor=north west, inner sep=0.75pt, align=left] {parameters};
    \draw (73,132) node [anchor=north west, inner sep=0.75pt, align=left] {cost};
    \draw (252,82) node [anchor=north west, inner sep=0.75pt, align=left] {action};
\end{tikzpicture}
    \caption{Diagram of the MPC-based RL architecture}
    \label{introduction:fig:mpcrl-diagram}
\end{figure}
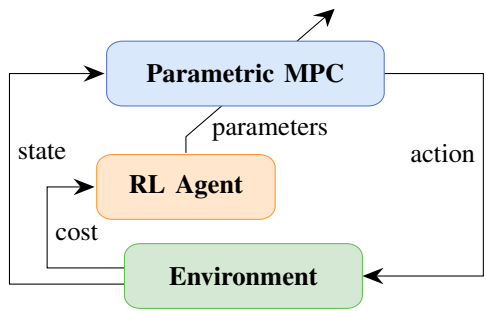

Recently, \cite{gros_2020_datadriven} proposed and justified the use of MPC as function approximation of the optimal policy in model-based RL. \cref{introduction:fig:mpcrl-diagram} depicts a schematic overview of this architecture. In such a scheme, the MPC controller's optimisation problem acts both as policy provider and value function approximation of the RL task. Concurrently, the learning algorithm (such as Q-learning and stochastic/deterministic policy gradient) is tasked with adjusting the parametrisation of the controller in an effort to directly or indirectly discover the optimal policy, thus improving closed-loop performance in a data-driven fashion. In this way, despite the presence of mismatches between the prediction model and the real system, the MPC control scheme is able to learn and deliver at convergence the optimal policy and value functions of the underlying RL task, granted the MPC parametrisation is rich enough. Unlike the more traditional pipeline, where a system identification phase (which selects model parameters that match the system's behaviour with observed data by, e.g., minimizing the predicted mean-squared-error) is followed by the controller design (possibly, iteratively), this approach directly tunes the parametrisation to maximize performance, irrespective of the controller's predictive accuracy. Such a rationale is often called Identification for Control, and advocates that the best parametrisation for control shall not yield the best predictions, but rather the best performance on the true system \cite{piga_2019_performance,gros_2020_datadriven,zanon_2021_safe}. In contrast to NN-based RL, the key advantages of the approach adopted in this paper are multiple:
\begin{itemize}
    \item The MPC scheme at its core allows to easily integrate prior information on the system in the form of an expert-based, possibly imperfect, prediction model.
    \item MPC-based agents are in general more amenable to analysis and certification in terms of stability and constraint satisfaction, a benefit supported by the extensive body of literature on MPC \cite{rawlings_2017_model}. Although challenging and beyond the scope of the proposed approach, it is possible to ensure that theoretical properties, such as stability, recursive feasibility and constraint guarantees, are preserved during the RL learning process \cite{zanon_2021_safe,gros_2022_learning}. On the other hand, NNs are notoriously black-box, lack interpretability, and often provide no guarantees prior to convergence.
    \item MPC controllers can also take constraints on states and actions into account in an explicit and structured way, which NNs are generally incapable of or can do poorly.
\end{itemize}
Compared to traditional non-learning MPC formulations, the proposed approach enjoys the following benefits:
\begin{itemize}
    \item The need for extensive open-loop data collection and system identification phase is eliminated, as the MPC parametrisation is automatically adjusted based on closed-loop data to improve performance, rather than prediction accuracy. This bypasses the issue of model inaccuracies and endows the scheme with online adaptation capabilities.
    \item By appropriately penalising constraint violations, the RL algorithm can guide the learning process to produce an MPC policy that satisfies system constraints.
\end{itemize}
All combined, these advantages can foster a safer training where the agent may enjoy a more stable and constraint-abiding learning process, opening up in the future the possibility of training directly in the real world \cite{dulac_2021_challenges}, and contributing to the suitability of this framework for the RM control problem. In particular, the proposed MPC-RL method automatically learns from data a policy that optimally balances congestion and queue lengths, without requiring further manual tuning. This bypasses the issue of model inaccuracies in traffic predictions, which are often the primary concern in non-myopic control strategies. It also embeds adaptivity in the RM strategy in the face of stochasticies (e.g., demand forecasts or the behaviour of drivers can vary significantly across different cities or countries, meaning that system identification performed in one region may not generalise well to another). Besides, we stress again that improving the accuracy of the prediction model may not necessarily result in a better closed-loop performance. Hence, this framework allows for improvements to the control policy without focusing on nor needing an accurate prediction model or identification as a separate step in the design process. Additionally, the approach strengthens constraint satisfaction, which is crucial in real-world applications like RM control to avoid critical congested scenarios.

This paper contributes to the state of the art by proposing a novel approach, which embeds the MPC framework within an RL algorithm, to tackle the highway RM control problem. Specifically:
\begin{enumerate}
    \item To the best of the authors’ knowledge, an MPC-based RL method inspired by \cite{gros_2020_datadriven} is proposed for the first time to solve the RM control problem. The method combines a differentiable parametrised MPC scheme, which acts as policy provider and function approximation, with an online second-order Q-learning algorithm. The Q-learning algorithm adjusts the MPC parametrisation via gradient updates based only on observed closed-loop data, allowing to automatically learn a congestion-free and constraint-abiding control policy. As explained earlier, this approach overcomes the issue of model inaccuracies and uncertainties in the MPC and in the context of RM.
    
    \item The proposed RM strategy is then implemented and tested on a benchmark highway stretch, where it is trained in an online fashion to simultaneously avoid bottlenecks and excessive queue lengths at the on-ramp, incorporating penalties for constraint violations directly into the MPC objective and the RL reward function. Simulation results underscore the effectiveness of the MPC-RL approach. Notably, starting from a poorly tuned controller with an imprecise prediction model, the proposed approach demonstrates quick learning in reducing congestion and satisfying the queue constraint. These results are further validated with a comparison with other state-of-the-art traffic controllers, both model-based and learning-based. Our method empirically outperforms them in performance and sample-efficiency.
\end{enumerate}

The paper is organised as follows. \Cref{sections:related_work} summarises relevant works in MPC, RL, and their combination in RM applications. Background on modelling and controlling traffic networks via MPC is provided in \Cref{sections:background}. \Cref{sections:methodology} presents the proposed methodology, showing how MPC can be combined with RL, as well as explaining the MPC-based RL algorithm itself. The method is then applied to a numerical case study in \Cref{sections:numerical}. Finally, \Cref{sections:conclusions} concludes the paper with some final remarks and future directions.

\section{Related Work} \label{sections:related_work}

Several methods have been proposed to address the RM problem \cite{siri_2021_freeway}. In this section, we report and discuss recent developments in model-based and model-free approaches, as well as the combination of the two, in comparison to our proposed methodology.  

\subsection{MPC Approaches for Ramp Metering Control}

A well-studied approach to tackle RM is MPC. First formalized in \cite{deschutter_1998_optimal}, it was deployed for RM \cite{bellemans_2006_model} and coordinated traffic control   \cite{hegyi_2002_optimal,hegyi_2004_model, hegyi_2005_model}, and is still subject of research, e.g., \cite{han_2020_hierarchical,todorovic_2022_distributed}. As explained in \Cref{sections:introduction}, its predictive capabilities make MPC highly desirable, but at the same time leave it susceptible to model mismatches and uncertainties. When these disturbances are assumed to be modelled, one can resort to robust or stochastic formulations to mitigate their impact, as proposed in, e.g., \cite{liu_2022_scenario}; however, such assumptions are in general hard to satisfy in practice, and the additional computational burden can be taxing. For these reasons, MPC techniques for traffic control that can adapt to and circumvent model mismatches and uncertainties are still a vibrant research area. 

\subsection{Model-free Approaches for Ramp Metering Control}

Perhaps the most established RM approach is ALINEA \cite{papageorgiou_1991_alinea}, a local feedback strategy that seeks to maximise freeway flow in a merging area by keeping the bottleneck density near critical value. Many extensions have been proposed to enhance its basic formulation, such as PI-ALINEA, which deals with distant downstream bottlenecks \cite{wang_2014_local}, and FF-ALINEA, which anticipates the evolution of a bottleneck density \cite{frejo_2019_feedforward}. However, these closed-loop controllers are fixed since they have no means to adapt their gains to changing conditions, and thus suffer from, e.g., uncertainties in traffic parameters and poor tuning. Ways have been devised to automatically account for it, e.g., via Kalman filtering to estimate the traffic parameters \cite{smaragdis_2004_flow} or based on historic data \cite{chen_2019_adaptive}. Unfortunately, these methods are still plagued with a fundamentally myopic control strategy, and either assume that parameters evolve according to a known observable model, or that a huge amount of offline recorded data is available and is representative of both current and future traffic scenarios. Moreover, these feedback methods, alongside MPC, often need to be paired with a mechanism for traffic state estimation, enabling them to respond to real-time traffic behaviour \cite{wang_2005_realtime}.

More recently, RL has gained popularity thanks to its ability to automatically learn an optimal policy solely by interacting with the environment, addressing the issue of uncertain and unknown dynamics. Earlier works adopted tabular Q-learning for the RM problem \cite{davarynejad_2011_motorway,fares_2014_freeway,ivanjko_2015_ramp}. However, with function approximations becoming more and more predominant, especially NNs, DRL algorithms have emerged as a relevant alternative to model-based controllers.
In \cite{wang_2022_integrated}, different policy gradient DRL algorithms are deployed in freeway control and compared. While results are promising, the issue of state constraints is not addressed, especially on the queue length, which should be contained to avoid spillback \cite{siri_2021_freeway}. This is a fundamental issue in RL, since these algorithms can often be made aware of the presence of constraints only when violations occur and are appropriately penalised in the reward, whereas MPC can handle constraints explicitly in its formulation. In \cite{han_2022_physics} it is proposed to use a combination of offline historic data and online synthetic data in an iterative way to avoid the RL model from being trapped in an inaccurate training environment. This work underscores the severe impact of high-quality data on the RL learning process, and raises the question of whether research efforts might be more effectively directed towards the development of online model-based learning algorithms that can learn directly from closed-loop interactions with the target environment while addressing constraints more explicitly. DRL has also facilitated the advancement of multi-agent solutions \cite{eltantawy_2013_multiagent}, which are especially useful in large-scale highway control problems. However, in addition to the aforementioned issues, multi-agent RL must also address the nonstationarity of its learning targets.

\subsection{Combination of MPC and RL for Ramp Metering} \label{sections:sub:mpc+rl}

The use of learning-based MPC schemes for traffic management has been investigated in the recent literature. For instance, \cite{abduljabbar_2021_development} shows how recurrent NNs can be used to predict traffic behaviour, and \cite{gu_2023_deep} proposes to adopt such a network as prediction model in their MPC scheme. 

Nonetheless, to the best of the authors' knowledge, the combination of MPC and RL in traffic control has received scarce attention, with \cite{remmerswaal_2022_combined,sun_2024_novel} the only relevant works. Therein, an MPC-DRL architecture is proposed that hierarchically combines the two controllers: the high-level MPC controller, which runs at a lower frequency, provides initial optimality while explicitly incorporating constraints, whereas the low-level DRL agent operates at a higher frequency and aims to compensate for model mismatches in the MPC. The MPC and RL control actions are then linearly combined together and fed to the freeway environment as a unique control input.

Given that the final control signal is a summation of an optimisation-based one and a learning-based one, it is anticipated that the resulting policy may not always be able to prevent constraint violations, particularly early in the training process. However, a benefit of \cite{sun_2024_novel} compared to the proposed methodology is the higher control frequency at which the DRL policy is run, granting quicker control response. This is especially advantageous in applications with fast dynamics, although traffic systems are typically slow, with sampling times up to tens of seconds. Note that the slower control frequency of our method is not inherently limiting as, with the improving hardware capabilities and faster solvers in recent years, increasing this frequency is feasible. The two methodologies also differ in computational complexity. While our approach is more demanding during the online training phase due to the need to solve the MPC problem twice per time step, it becomes more efficient at deployment, as it requires only one MPC calculation per time step \cite{gros_2020_datadriven}. In contrast, \cite{sun_2024_novel} runs both the MPC and DRL policies at every step, leading to higher computational costs at inference. Moreover, note that if training is performed offline, e.g., via an off-policy algorithm, then the additional computational burden of our method becomes irrelevant to online performance.

\subsection{Comparison with Proposed Methodology}

Compared to feedback strategies such as ALINEA, the MPC-based RL framework \cite{gros_2020_datadriven} we propose to use for RM leverages as its core an MPC controller, which, thanks to its predictive capabilities, can yield non-myopic policies. At the same time, its RL module allows to automatically adjust the MPC controller, and thus, it circumvents some of the challenges that conventional MPC typically struggles to address, i.e., model misspecifications and poor identification and tuning. This framework also boasts a higher degree of interpretability, better handling of constraints and integration of prior knowledge compared to other model-free RL methods. Furthermore, in contrast to \cite{sun_2024_novel}, only one policy provider (the differentiable MPC controller itself) is present. For this reason, as remarked in \Cref{sections:introduction}, this approach has the potential to result in fewer or no constraint violations at all depending on, e.g., the complexity of the underlying model and objective and the presence or lack of uncertainties \cite{zanon_2021_safe,gros_2022_learning}. Likewise, at deployment, the proposed method relies only on the learnt MPC problem, whereas the method from \cite{sun_2024_novel} must additionally compute the output from the NN actor.

\section{Background} \label{sections:background}

Three preliminary components are necessary to illustrate the proposed MPC-RL control method. First, a modelling framework must be employed to build a representation of the dynamical behaviour of the highway network, and how RM can influence it to prevent, e.g., bottlenecks and spill-backs. Then, we report a classical formulation of MPC for this task that makes use of such a model, and propose some modifications to it that are relevant to this work. Lastly, the essential concepts of RL are briefly introduced.

\subsubsection*{Notation}
The set of indices $\{1,\ldots,N\} \subseteq \mathbb{N}$ is denoted as $\mathcal{N}$. Vector quantities are in bold. Inequalities on vectors are applied element-wise. To avoid confusion when other subscripts are present, dynamic quantities (e.g., $\rho$) at time step $k$ can be also referred to as, e.g., $\rho_j(k)$. The $i$-th index of predicted quantities based on the information available at time step $k$ are denoted as, e.g., $\hat{y}_{i|k}$.

\subsection{METANET Modelling Framework} \label{sections:sub:metanet}

Based on the traffic phenomena to investigate, different approaches to modelling a highway network exist \cite{treiber_2013_traffic}. In this paper, the macroscopic second-order METANET framework is employed to obtain a discrete-time dynamical representation of the behaviour of highway traffic under RM control. Proposed as a simulation tool in \cite{papageorgiou_1989_macroscopic,messner_1990_metanet}, it made one of its first appearances in combination with MPC in \cite{bellemans_2006_model,hegyi_2005_model}, and has since been extended to cover several phenomena and traffic measures \cite{papageorgiou_2010_traffic,dabiri_2017_distributed}.

In the remainder of this section, the basics of METANET are introduced. In this framework, a network is represented as a directed graph where each segment (a continuous stretch of highway with homogeneous properties) is modelled as an arc, and origins are represented as nodes. Virtual nodes are also employed to separate segments with heterogeneous properties (e.g., due to a lane drop). We assume the network consists of $\card*{\mathcal{S}}$ segments and $\card*{\mathcal{O}}$ origins, where $\mathcal{S}$ and $\mathcal{O}$ are the sets of indices of segments and origins, respectively. At time step $k$, segment $j \in \mathcal{S}$ is characterized by its traffic density $\rho_j(k)$ and a mean speed $v_j(k)$, which form its state. Conversely, origin $o \in \mathcal{O}$ is characterized by the queue length $w_o(k)$. In the destination-independent variant, the evolution in time of segment states is described by the following equations:
\begin{align}
    \rho_j(k + 1) ={} & \rho_j(k) + \frac{T}{L_j \lambda_j} \bigl(q_{j-1}(k) - q_j(k) + r_o(k)\bigr),  \\
    \begin{split}
        v_j(k + 1) ={} & v_j(k) + \frac{T}{\tau} \Bigl( V_\text{e}\bigl(\rho_j(k)\bigr) - v_j(k) \Bigr) \\
        & + \frac{T}{L_j} v_j(k) \bigr( v_{j-1}(k) - v_j(k) \bigl) \\
        & - \frac{\eta T}{\tau L_j} \frac{\rho_{j+1}(k) - \rho_j(k)}{\rho_j(k) + \kappa} \\
        & - \frac{ \mu T r_o(k) v_j(k) }{ L_j \lambda_j \bigr( \rho_j(k) + \kappa \bigl) },
    \end{split}  \\
    q_j(k) ={} & \lambda_j \rho_j(k) v_j(k) \label{background:eq:traffic-flow},  \\
    V_\text{e} \bigl(\rho(k)\bigr) \coloneqq{} & v_\text{free} \exp{\left( -\frac{1}{a} \left( \frac{\rho(k)}{\rho_\text{crit}} \right)^a \right)} \label{background:eq:equilibrium-speed},
\end{align}
where $T$ is the sampling time, $L_j$ and $\lambda_j$ are respectively the length of the segment and its number of lanes, \eqref{background:eq:traffic-flow} describes the traffic flow $q_j(k)$ of segment $j$ at time step $k$, \eqref{background:eq:equilibrium-speed} is the well-known equilibrium speed formula, and $\tau, \eta, \kappa, \mu, a, \rho_\text{crit}, v_\text{free}$ are model parameters describing different quantities and phenomena. In particular, $\rho_\text{crit}$ indicates the traffic density at which traffic flow transitions from free-flow conditions to congested conditions. At this density, the flow rate of vehicles through the network is at its maximum, and any increase in the number of vehicles beyond this point results in reduced speeds and increased congestion. The free-flow speed $v_\text{free}$ describes instead the speed at which vehicles can travel when the road is not congested or disrupted, i.e., it is the maximum speed that vehicles can achieve on a particular segment under low traffic volumes. Lastly, $r_o(k)$ is the incoming flow generated by origin $o$ connected to segment $j$; if none is connected, it is null. In these equations, the index $j-1$ refers to the segment upstream of $j$, and $j+1$ to the segment downstream. For origins, the queue length evolves as
\begin{equation} \label{background:eq:queue-dynamics}
    w_o(k + 1) = w_o(k) + T \bigl( d^w_o(k) - r_o(k) \bigr),
\end{equation}
where the origin's demand $d^w_o(k)$ acts as an uncontrollable external input affecting the queue dynamics. This work is mainly concerned with on-ramps; however, different types of models for origins exist, and interested readers are referred to \cite{papageorgiou_2010_traffic}. Having defined the queue length as in \eqref{background:eq:queue-dynamics}, the on-ramp flow can be computed as
{\allowdisplaybreaks
\begin{multline} \label{background:eq:ramp-flow}
    r_o(k) = \tilde{r}_o(k) \min \biggl\{ \\
    d^w_o(k) + \frac{w_o(k)}{T},
    C_o,
    C_o \left( \frac{\rho_\text{max}-\rho_{j_o}(k)}{ \rho_\text{max}-\rho_\text{crit} } \right) \biggr\},
\end{multline}}
where $C_o$ is the capacity of the origin, $\rho_{j_0}(k)$ is the density of the segment $j_o$ that is connected to this ramp, and $\rho_\text{max}$ is the maximum density segments can withstand (assumed here independent of index $j_o$, for sake of simplicity). Lastly, $\tilde{r}_o(k) \in [0,1]$ is the metering rate control action, which allows modulating the entering ramp flow to prevent, e.g., congestion and spillback. However, to avoid numerical issues in the MPC gradient-based solvers due to the $\min$ operator in \eqref{background:eq:ramp-flow}, which can cause the gradient to be zero over a vast region of the state-action space, it is more efficient to control the ramp flow $r_o$ directly. This requires imposing the additional terms in \eqref{background:eq:ramp-flow} as constraints on the control action, as shown in the next subsection.

Finally, the downstream boundary conditions are here considered to be affected by congestion, i.e., the highway segment $j_i$ that ends in destination $i \in \mathcal{D}$ is subject to the (virtual) downstream density modelled in \cite{hegyi_2004_model} as
\begin{equation}
    \rho_{j_i+1}(k) = \max{\Bigl\{ \min{\bigl\{ \rho_{j_i}(k), \rho_\text{crit} \bigr\}}, d^\rho_i\left(k\right) \Bigr\}},
\end{equation}
where $\mathcal{D}$ is the set of destination indices, $d^\rho_i(k)$ models the congestion density of destination $i$, and is regarded as another external disturbance.

In a more concise formulation, the states, actions and external factors at time $k$ can be lumped in vectors as
\begin{align}
    \begin{split} \label{background:eq:state-definition}
        \bm{x}_k ={} & \bigl[ \rho_1(k) \ \ \ldots \ \ \rho_{\card*{\mathcal{S}}}(k) \ \ v_1(k) \ \ \ldots \ \ v_{\card*{\mathcal{S}}}(k)  \\
            & \phantom{============} w_1(k) \ \ \ldots \ \ w_{\card*{\mathcal{O}}}(k) \bigr]^\top \hspace{2pt},
    \end{split}  \\
    \bm{u}_k ={} & \begin{bmatrix} r_1(k) & \ldots & r_{\card*{\mathcal{O}}}(k) \end{bmatrix}^\top, \label{background:eq:action-definition} \\
    \bm{d}_k ={} & \begin{bmatrix} d^w_1(k) & \ldots & d^w_{\card*{\mathcal{O}}}(k) & d^\rho_1(k) & \ldots & d^\rho_{\card*{\mathcal{D}}}(k) \end{bmatrix}^\top,
\end{align}
respectively. Then, the METANET framework can be exploited to build the nonlinear dynamics
\begin{equation} \label{background:eq:metanet-dynamics}
    \bm{x}_{k+1} = f(\bm{x}_k, \bm{u}_k, \bm{d}_k),
\end{equation}
where $f : \mathbb{R}^{n_x} \times \mathbb{R}^{n_u} \times \mathbb{R}^{n_d} \rightarrow \mathbb{R}^{n_x}$, with $n_x = 2\card*{\mathcal{S}} + \card*{\mathcal{O}}$, $n_u = \card*{\mathcal{O}}$, $n_d = \card*{\mathcal{O}} + \card*{\mathcal{D}}$. The next section details how these dynamics can be used in designing a model-based controller.

\subsection{MPC for Ramp Metering Control}

In what follows, the classical MPC-based paradigm for the RM control problem is presented and summarised.

MPC \cite{rawlings_2017_model} is a well-known control framework for dynamical systems that is based on the formulation of a finite-horizon optimal control problem, consisting of three fundamental pillars: the prediction of the evolution of the system's dynamics over a finite window, a suitable objective function to be minimised along this prediction window, and the inclusion of constraints on state and/or action variables. The first is typically achieved by means of a (often approximate) model of the system's behaviour, which is used to simulate the evolution of the states along the window. The second is a function that quantifies the performance of the controller and is often a trade-off between conflicting objectives. Thirdly, MPC allows to systematically integrate into its structure constraints on the states and/or actions, thus yielding policies that are able to take these constraints into account in an optimal way. On the whole, the framework offers quite a degree of flexibility, and it is unsurprising that it lends itself also to RM control problems.

We define the MPC prediction horizon $N_\text{p}$ and the control horizon $N_\text{c} \leq N_\text{p} / M$, where $M$ is a natural number to distinguish the process timescale $T$ from the controller timescale $T_\text{c} = T / M$. In other words, the process runs $M$ times faster than the controller, with the MPC being solved only every $M$ time steps (instead of at each time step). Once solved, the first optimal action is then applied for the next $M$ time steps, in a receding horizon fashion. Therefore, in the MPC controller timescale, control action indices are indicated as $i_\text{c}$ and can be related to indices $i$ of the dynamics timescale as $i_\text{c}(i) = \min{\left\{\left\lfloor i / M \right\rfloor, N_\text{c} - 1 \right\}}$.\footnote{This definition of $i_\text{c}(i)$ entails that the last action is kept constant for the remainder of the prediction horizon, a common choice in literature, though more complex solutions could be employed.} Additionally, we define a constrained RM control scenario by imposing at time step $k$, for the on-ramps in the network, the constraint
\begin{equation} \label{background:eq:queue-constraints}
    w_o(k) \le w_\text{max}, \quad \forall o \in \mathcal{O},
\end{equation}
where $w_\text{max}$ represents the maximum queue length the ramps should experience. For the sake of simplicity, it is assumed here that $w_\text{max}$ is independent of the index $o$, and that all queue lengths are subject to this constraint, though it could be readily applied to only a subset of them. As mentioned in \Cref{sections:related_work}, this is usually done to prevent spillback, and similarly constrained scenarios can be found in, e.g., \cite{hegyi_2005_model}. This constraint is not to be considered hard, but we favour control strategies that are able to satisfy it, if not always, at least for most time steps.

Altogether, given the current state $x_k$ of the network, the MPC controller is given by
\begin{mini!}
    { \scriptstyle \hat{\bm{X}}_k, \hat{\bm{U}}_k, \bm\Sigma }{
        \sum_{i=0}^{N_\text{p}}{L_\text{T}(\hat{\bm{x}}_{i|k})}
        + \xi \sum_{i=0}^{N_\text{c} - 1}{L_\text{V}(\hat{\bm{u}}_{i|k})}
        + \sum_{i=0}^{N_\text{p}}{ \bm\zeta_i^\top \bm\sigma_i }
    }{ \label{background:eq:mpc-rm:scheme} }{ \label{background:eq:mpc-rm:objective} }
    \addConstraint{ \hat{\bm{x}}_{0|k} }{ = \bm{x}_k, }{}{ \label{background:eq:mpc-rm:initial-state} }
    \addConstraint{ \hat{\bm{x}}_{i+1|k} }{ 
        = f(\hat{\bm{x}}_{i|k}, \hat{\bm{u}}_{i_\text{c}(i)|k}, \bm{d}_k), \ }{ i = 0, \ldots, N_\text{p} - 1, 
    }{ \label{background:eq:mpc-rm:dynamics} }
    \addConstraint{ h(\hat{\bm{x}}_{i|k}, \hat{\bm{u}}_{i_\text{c}(i)|k}) }{ \leq 0, }{ i = 0, \ldots, N_\text{p}, }{ \label{background:eq:mpc-rm:ineq-constraints:h} }
    \addConstraint{ g(\hat{\bm{x}}_{i|k}, \bm\sigma_i) }{ \leq 0, }{ i = 0, \ldots, N_\text{p}, }{ \label{background:eq:mpc-rm:ineq-constraints:g} }
    \addConstraint{ \bm\sigma_i }{ \ge 0, }{ i = 0, \ldots, N_\text{p}, }{}
\end{mini!}
where the optimisation variables are the actions and states along the MPC control and prediction horizons, and the slack variables (whose purpose is explained later in the section), respectively in order
{\allowdisplaybreaks
\begin{align}
    \hat{\bm{X}}_k &{}= \begin{bmatrix} 
        \hat{\bm{x}}_{0|k}^\top & \ldots & \hat{\bm{x}}_{N_\text{p}|k}^\top 
    \end{bmatrix}^\top
    \in \mathbb{R}^{ n_x \left( N_\text{p} + 1 \right) },  \\ 
    \hat{\bm{U}}_k &{}= \begin{bmatrix} 
        \hat{\bm{u}}_{0|k}^\top & \ldots & \hat{\bm{u}}_{N_\text{c} - 1|k}^\top 
    \end{bmatrix}^\top
    \in \mathbb{R}^{ n_u N_\text{c} },  \\
    \bm\Sigma &{}= \begin{bmatrix} 
        \bm\sigma_0^\top & \ldots & \bm\sigma_{N_\text{p}}^\top 
    \end{bmatrix}^\top
    \in \mathbb{R}^{ \card*{\mathcal{O}} \left( N_\text{p} + 1 \right) },
\end{align}}
where $\hat{\bm{X}}_k$ and $\hat{\bm{U}}_k$ are defined in analogy to \eqref{background:eq:state-definition} and \eqref{background:eq:action-definition} respectively, and $\bm\sigma_i$ collects the slack variables at the predicted step $i$ corresponding to all on-ramps, i.e., $\bm\sigma_i = \begin{bmatrix} \sigma_i^1 & \ldots & \sigma_i^{\card*{\mathcal{O}}} \end{bmatrix}^\top$. As aforementioned, once this optimisation problem is solved, we apply the optimal control action $\hat{\bm{u}}^\star_{0|k}$ from time step $k$ to $k + M - 1$, as per the receding horizon approach.

The objective \eqref{background:eq:mpc-rm:objective} comprises three terms. The first is the total time spent (TTS), a metric frequently used to quantify efficiency of traffic control, defined as $L_\text{T} : \mathbb{R}^{n_x} \rightarrow \mathbb{R}$
\begin{equation}
    L_\text{T}(\bm{x}_k) \coloneqq T \left(\sum_{j \in \mathcal{S}}{L_j \lambda_j \rho_j(k)} + \sum_{o \in \mathcal{O}}{w_o(k)}\right).
\end{equation}
The second term of \eqref{background:eq:mpc-rm:objective} is a cost proportional to the variability of the inputs along the control horizon, where $L_\text{V} : \mathbb{R}^{n_u} \rightarrow \mathbb{R}$ is
\begin{equation} \label{background:eq:mpc-rm:objective:var}
    L_\text{V}(\bm{u}_k) \coloneqq \sum_{o \in \mathcal{O}}{ \left(\frac{r_o(k) - r_o(k - 1)}{C_o}\right)^2 }.
\end{equation}
Lastly, the third term in \eqref{background:eq:mpc-rm:objective} acts as a penalty for the slack variables. The nonnegative weights $\xi \in \mathbb{R}$ and $\bm\zeta_i \in \mathbb{R}^{ \card*{\mathcal{O}}}$, $i=0,\ldots,N_\text{p}$, govern the trade-off between these three terms and are selected by the designer so as to encode the desired behaviour of the controller \cite{hegyi_2002_optimal}. \Cref{sections:methodology} will show how these parameters (under the new symbols $\theta_\text{V}$ and $\bm\Theta_\text{C}$), together with others, can be adjusted automatically via RL instead of having to be specified manually.

In \eqref{background:eq:mpc-rm:ineq-constraints:h} and \eqref{background:eq:mpc-rm:ineq-constraints:g}, $h : \mathbb{R}^{n_x} \times \mathbb{R}^{n_u} \rightarrow \mathbb{R}^{4 \card*{\mathcal{O}}}$ and $g : \mathbb{R}^{n_x} \times \mathbb{R}^{\card*{\mathcal{O}}} \rightarrow \mathbb{R}^{\card*{\mathcal{O}}}$ correspond to all the inequality constraints to be imposed in the MPC problem. In particular, they are defined as 
\begin{align}
    h(\bm{x}_i, \bm{u}_{i_\text{c}}) \coloneqq{}& \begin{bmatrix}
        h_1(\bm{x}_i, \bm{u}_{i_\text{c}})^\top &
        \ldots &
        h_{\card*{\mathcal{O}}}(\bm{x}_i, \bm{u}_{i_\text{c}})^\top
    \end{bmatrix}^\top, \\
    g(\bm{x}_i, \bm\sigma_i) \coloneqq{}& \begin{bmatrix}
        g_1(\bm{x}_i, \bm\sigma_i) &
        \ldots &
        g_{\card*{\mathcal{O}}}(\bm{x}_i, \bm\sigma_i)
    \end{bmatrix}^\top,
\end{align}
where
\begin{align}
    h_o(\bm{x}_i, \bm{u}_{i_\text{c}}) &\coloneqq \begin{bmatrix} 
        - r_o(i_\text{c}) \\
        r_o(i_\text{c}) - C_o \\
        r_o(i_\text{c}) - d^w_o(i) - \frac{w_o(i)}{T} \\
        r_o(i_\text{c}) - C_o \left(\frac{\rho_\text{max} - \rho_{j_o}(i)}{\rho_\text{max} - \rho_\text{crit}}\right)
    \end{bmatrix}, &\forall o \in \mathcal{O}, 
    \label{background:eq:equivalent-flow-constraints} \\
    g_o(\bm{x}_i, \bm\sigma_i) &\coloneqq w_o(i) - w_\text{max} - \sigma_i^o, &\forall o \in \mathcal{O}, 
    \label{background:eq:soft-queue-constraints}
\end{align}
and we have dropped the dependency on $i$ in $i_\text{c}$ for readability. The constraints $h_o(\bm{x}_i, \bm{u}_{i_\text{c}}) \leq 0$ emulate the behaviour of \eqref{background:eq:ramp-flow} and prevent the control action from assuming invalid values. The constraints $g_o(\bm{x}_i, \bm\sigma_i) \leq 0$ instead incorporate the queue requirements \eqref{background:eq:queue-constraints} into the MPC controller as soft constraints, where $\bm\sigma_k \ge 0$ is a slack variable whose purpose is to relax the constraint in order to avoid infeasibility issues, and so it must be appropriately penalised in the objective \eqref{background:eq:mpc-rm:objective}. We opt for a soft constraint because
\begin{enumerate*}[label=\arabic*)]  
    \item it is assumed that small violations to this constraint, although undesired, are tolerable, and
    \item we need a mechanism to both relax constraints during the learning process and penalise violations in order to inform the agent about policies that violate constraints and ones that do not.
\end{enumerate*}

Note that in \eqref{background:eq:mpc-rm:initial-state} we assume that the current process state $\bm{x}_k$ is fully measurable. If this is not the case, an observer is required to provide an estimate of it. Likewise, in \eqref{background:eq:mpc-rm:dynamics} we assume the external inputs $\bm{d}_k$ to be fully known. If this is not the case, a forecast of the evolution of these quantities along the MPC horizon is required. 

\subsection{Reinforcement Learning}

To better understand how MPC can be integrated with RL, a brief introduction to the latter is here provided. In what follows, we abide by the canonical RL approach \cite{sutton_2018_reinforcement}. Given the continuous state $\bm{s}$ and action $\bm{a}$, a discrete-time Markov Decision Process with state transitions $\bm{s} \xrightarrow{\bm{a}} \bm{s}_+$ and underlying conditional probability density
\begin{equation}
    \probab*{\bm{s}_+ \given \bm{s}, \bm{a}} : \mathbb{R}^{n_s} \times \mathbb{R}^{n_s} \times \mathbb{R}^{n_a} \rightarrow \left[0, 1\right]
\end{equation}
is used to model the discrete-time system dynamics. The performance of a given deterministic policy $\pi_{\bm\theta} : \mathbb{R}^{n_s} \rightarrow \mathbb{R}^{n_a}$ parametrized in $\bm\theta \in \mathbb{R}^{n_{\bm\theta}}$ is defined as
\begin{equation} \label{background:eq:rl:performance}
    J(\pi_{\bm\theta}) \coloneqq \expval*{ \sum_{k=0}^{\infty}{\gamma^k L \bigl(\bm{s}_k, \pi_{\bm\theta}(\bm{s}_k)\bigr)} },
\end{equation}
where $L : \mathbb{R}^{n_s} \times \mathbb{R}^{n_a} \rightarrow \mathbb{R}$ is a stage cost function and $\gamma \in (0, 1]$ the discount factor. The goal of the RL algorithm is then to find the optimal policy $\pi^\star_{\bm\theta}$ as
\begin{argmini}
    { \scriptstyle \bm\theta }{ J(\pi_{\bm\theta})}{ \label{background:eq:rl:optimal_policy} }{ \pi_{\bm\theta}^\star = }.
\end{argmini}
Since it is in general impossible to find and characterise the true unknown optimal value functions and policy $V_\star$, $Q_\star$, $\pi_\star$, function approximation techniques (such as NN and, as in this paper, MPC) have been employed as a powerful alternative for tackling problem \eqref{background:eq:rl:optimal_policy} \cite{busoniu_2017_reinforcement}. Depending on the algorithm, these allow formulating the approximations $V_{\bm\theta}$, $Q_{\bm\theta}$, and $\pi_{\bm\theta}$, which are then used to solve \eqref{background:eq:rl:optimal_policy} either directly or indirectly via iterative gradient updates of the parametrisation
\begin{equation}  \label{background:eq:rl:gradient_update}
    \bm\theta \leftarrow \bm\theta 
    - \alpha \nabla_{\bm\theta} \sum_{i=1}^{m}{ 
        \psi \left( \bm{s}_i, \bm{a}_i, \bm{s}_{i+1}, \bm\theta \right) 
    },
\end{equation}
where $\alpha \in \mathbb{R}_+$ is the learning rate, $m$ is the size of the batch of observations used in the update, and $\psi$ captures the controller's performance and varies from algorithm to algorithm. Direct (or policy-based) methods explicitly parametrise and learn the approximation $\pi_{\bm\theta}$ of the optimal policy itself. For instance, policy gradient methods attempt to solve \eqref{background:eq:rl:optimal_policy} directly by moving along the gradient of the policy, i.e.,
\begin{equation}
    \expval*{ \psi \left(\bm{s}_i, \bm{a}_i, \bm{s}_{i+1}, \bm\theta\right) } = J(\pi_{\bm\theta}).
\end{equation}
On the other hand, indirect (or value-based) methods opt to indirectly derive the optimal policy by estimating and learning the value functions $V_{\bm\theta},Q_{\bm\theta}$ from the underlying RL task, and implicitly derive the policy from these. A member of this category, Q-learning learns to approximate the action-value by solving
\begin{mini}
    { \scriptstyle \bm\theta }{ \expval*{\norm*{Q_\star(s,a) - Q_{\bm\theta}(s,a)}^2} }{ \label{background:eq:rl:bellman-residual-problem} }{},
\end{mini}
with the hope of indirectly recovering the optimal policy from it. In its first-order recursive (i.e., with $m=1$) formulation, $\bm\theta$ is updated via \eqref{background:eq:rl:gradient_update} with $\psi(\bm{s}_i,\bm{a}_i,\bm{s}_{i+1},\bm\theta) = \delta_i^2$, where $\delta_i$ is the well-known temporal difference (TD) error:
\begin{equation} \label{background:eq:rl:td-error}
    \delta_i = L(\bm{s}_i, \bm{a}_i) + \gamma V_{\bm\theta}(\bm{s}_{i+1}) - Q_{\bm\theta}(\bm{s}_i, \bm{a}_i).
\end{equation}

\section{MPC-based RL for Ramp Metering} \label{sections:methodology}

As discussed in \Cref{sections:introduction}, the closed-loop performance of the MPC controller \eqref{background:eq:mpc-rm:scheme} is dependent on its components, especially the prediction model. Mismatches in the dynamics can undermine the ability of MPC to reliably predict the system behaviour, thus impacting the controller performance and its ability to avoid constraint violations. The highly nonlinear nature of the METANET model contributes to making this issue even more relevant, since the model calibration requires the application of some nonlinear technique, e.g., nonlinear least-squares \cite{ngoduy_2003_automated}. In what follows, we show how these shortcomings can be addressed by leveraging online closed-loop data to learn most of the MPC parameters automatically via RL.

\subsection{Ramp Metering as an RL Task}

To appropriately apply an RL algorithm to the RM task, we first need to define a proper stage cost $L(x_k, u_k)$ to quantify the performance of a policy in controlling the metering, as per \eqref{background:eq:rl:performance}. Also referred to as reward in literature (i.e., the negative of the cost), it defines the overall return that the RL algorithm must minimise and it encodes the objective that the designer wants to see minimised by the MPC control policy. Consider
\begin{equation} \label{methodology:eq:rl-stage-cost}
    L(\bm{x}_k, \bm{u}_k) \coloneqq c_\text{T} L_\text{T}(\bm{x}_k) + c_\text{V} L_\text{V}(\bm{u}_k) + c_\text{C} L_\text{C}(\bm{x}_k),
\end{equation}
where $L_\text{C} : \mathbb{R}^{n_x} \rightarrow \mathbb{R}$ is
\begin{equation}
    L_\text{C}(\bm{x}_k) \coloneqq \sum_{o \in \mathcal{O}} \max{\bigl\{0, w_o(k) - w_\text{max}\bigr\}}.
\end{equation}
The first term in \eqref{methodology:eq:rl-stage-cost} penalises the time spent travelling through the network and waiting in queues, at the current time step $k$. This term is representative of the TTS criterion, which is often the primary target of RM control strategies in literature. The second term penalises variations of the current control action $r_o(k)$ compared to the previous instant $r_o(k - 1)$, and attempts to punish policies that are too jerky in the input signal. Lastly, the third contribution acts as the RL counterpart to the MPC soft constraint \eqref{background:eq:soft-queue-constraints} and penalisation of the corresponding slack variables in the MPC objective\eqref{background:eq:mpc-rm:objective}: it penalises violations of the queue constraint so that the agent is encouraged to come up with policies that are able to satisfy this constraint. Scalars $c_\text{T}$, $c_\text{V}$, and $c_\text{C}$ are weights that balance each term differently, and it is the designer's task to choose these in such a way to properly embed the learning objective in the RL stage cost.\footnote{In our work, the values for $c_\text{T}$, $c_\text{V}$, and $c_\text{C}$ were manually adjusted to achieve a reasonable learning process.} Next, we delve into how MPC can be used as function approximation to solve the ramp metering RL task.

\subsection{MPC as Function Approximation in RL} \label{sections:sub:mpc-func-approximator}

While the RL stage cost \eqref{methodology:eq:rl-stage-cost} defines and guides the learning process, a function approximation is needed to estimate the underlying optimal value and policy functions. A recurrent candidate for this rule in the literature is NNs. However, in \Cref{sections:introduction} it was discussed how \cite{gros_2020_datadriven} suggests the employment of a parametric MPC scheme as function approximation in the RL problem, and lies the foundations of its theory. In this paper, we follow this paradigm. In particular, we choose to parameterise the cost function, the prediction model as well as some of the constraints in \eqref{background:eq:mpc-rm:scheme}. Then, the RL algorithm is applied to adjust the parametrisation based on observed state transitions and rewards, in order to achieve better closed-loop performance in terms of \eqref{background:eq:rl:performance}.

Consider the MPC controller with parametrisation $\bm\theta$
\begin{mini!}
    { \scriptstyle \hat{\bm{X}}_k, \hat{\bm{U}}_k, \bm\Sigma }{
        \sum_{i=0}^{N_\text{p}}{
            \gamma^i \Bigl(\theta_\text{T} L_\text{T}(\hat{\bm{x}}_{i|k})
            + {\bm\theta_\text{C}^i}^\top \bm\sigma_i
        \Bigr)}
        \notag
    }{ \label{methodology:eq:mpc-func-approx:scheme} }{}
    \breakObjective{
        + \theta_\text{V} \sum_{i=0}^{N_\text{c} - 1}{\gamma^{i M} L_\text{V}(\hat{\bm{u}}_{i|k})}
        + \lambda_{\bm\theta}(\hat{\bm{x}}_{0|k})
        \notag
    }
    \breakObjective{
        + \sum_{i=1}^{N_\text{p} - 1}{\gamma^i \ell_{\bm\theta}(\bm{x}_k)}
        + \gamma^{N_\text{p}} \Gamma_{\bm\theta}(\bm{x}_{N_\text{p}})
        \label{methodology:eq:mpc-func-approx:obj}
    }
    \addConstraint{ \hat{\bm{x}}_{0|k} }{ = \bm{x}_k }{ \label{methodology:eq:mpc-func-approx:init-state} }
    \addConstraint{ \hat{\bm{x}}_{i+1|k} }{ 
        = f_{\bm\theta}(\hat{\bm{x}}_{i|k}, \hat{\bm{u}}_{i_\text{c}(i)|k}, \bm{d}_k), 
        i = 0, \ldots, N_\text{p} - 1, 
    }{ \label{methodology:eq:mpc-func-approx:dynamics} }
    \addConstraint{ h_{\bm\theta}(\hat{\bm{x}}_{i|k}, \hat{\bm{u}}_{i_\text{c}(i)|k}) }{ 
        \leq 0, \hspace{36.25pt} i = 0, \ldots, N_\text{p}, 
    }{ \label{methodology:eq:mpc-func-approx:h} }
    \addConstraint{ g(\hat{\bm{x}}_{i|k}, \bm\sigma_i) }{ 
        \leq 0, \hspace{61.75pt} i = 0, \ldots, N_\text{p}, 
    }{ \label{methodology:eq:mpc-func-approx:g} }
    \addConstraint{ \bm\sigma_i }{ 
        \ge 0, \hspace{95.75pt} i = 0, \ldots, N_\text{p}. 
    }{ \label{methodology:eq:mpc-func-approx:sigma-constraint} }
\end{mini!}
with $\bm\Theta_\text{C} = \begin{bmatrix} {\bm\theta_\text{C}^0}^\top & \ldots & {\bm\theta_\text{C}^{N_\text{p}}}^\top \end{bmatrix}^\top \in \mathbb{R}^{\card*{\mathcal{O}} \left(N_\text{p}+1\right)}$. Because this controller acts as function approximation and must be able to accurately predict the realisations of the RL stage cost along the prediction horizon, the objective \eqref{methodology:eq:mpc-func-approx:obj} is devised so as to mimic \eqref{methodology:eq:rl-stage-cost}. Furthermore, note the additional parameterised terms $\lambda_{\bm\theta}, \ell_{\bm\theta}, \Gamma_{\bm\theta} : \mathbb{R}^{n_x} \rightarrow \mathbb{R}$. These are the learnable initial, stage, and terminal costs respectively, and serve the purpose of enriching the parametrisation and facilitating generalisation across the state-action space. The initial cost is a linear combination of the state
\begin{equation}
    \lambda_{\bm\theta}(\bm{x}_k) \coloneqq \sum_{j \in \mathcal{S}}{\left(
        \theta^\rho_{\lambda,j} \frac{\rho_j(k)}{\rho_\text{max}}
        + \theta^v_{\lambda,j} \frac{v_j(k)}{v_\text{max}}
        \right)}
    + \sum_{o \in \mathcal{O}}{\theta^w_{\lambda,o} \frac{w_o(k)}{w_\text{max}}},
\end{equation}
where $\rho_\text{max}$, $v_\text{max}$, $w_\text{max}$ are here used as fixed, non-learnable normalization constants in order to improve the stability of the learning process. As explained in \cite{gros_2020_datadriven}, since our objective is economic, this initial cost is required in order for the MPC scheme to recover the optimal policy. Conversely, stage and terminal costs are chosen to be quadratic quadratic
\begin{align}
    \begin{split} \label{methodology:eq:mpc-func-approx:stage-cost}
        \ell_{\bm\theta}(\bm{x}_k) \coloneqq {} &
        \sum_{j \in \mathcal{S}}{
            \theta^\rho_{\ell,j} \left(\frac{\rho_j(k) - \rho_\text{sp}}{\rho_\text{max}}\right)^2
        } \\
        & + \sum_{j \in \mathcal{S}}{
        \theta^v_{\ell,j} \left(\frac{v_j(k) - v_\text{sp}}{v_\text{max}}\right)^2
        } \\
        & + \sum_{o \in \mathcal{O}}{\theta^w_{\ell,o} \left(\frac{ w_o(k)}{w_\text{max}}\right)^2},
    \end{split} \\
    \begin{split} \label{methodology:eq:mpc-func-approx:terminal-cost}
        \Gamma_{\bm\theta}(\bm{x}_k) \coloneqq {} &
        \sum_{j \in \mathcal{S}}{
            \theta^\rho_{\Gamma_,j} \left(\frac{\rho_j(k) - \rho_\text{sp}}{\rho_\text{max}}\right)^2
        } \\
        & + \sum_{j \in \mathcal{S}}{
        \theta^v_{\Gamma_,j} \left(\frac{v_j(k) - v_\text{sp}}{v_\text{max}}\right)^2
        } \\
        & + \sum_{o \in \mathcal{O}}{\theta^w_{\Gamma_,o} \left(\frac{ w_o(k)}{w_\text{max}}\right)^2},
    \end{split}
\end{align}
where densities and speeds are encouraged to track the fixed non-learnable setpoints $\rho_\text{sp}$ and $v_\text{sp}$ respectively, and the queues to be as small as possible. Regarding the METANET model $f_{\bm\theta}$ in \eqref{methodology:eq:mpc-func-approx:dynamics}, we allow the RL algorithm to adjust two fundamental parameters in the dynamics, i.e., $\tilde{\rho}_\text{crit}$ and $\tilde{a}$ (which in general can differ, during the learning process, from their counterparts $\rho_\text{crit}$, $a$ of the real dynamics). In this way, the agent is able to partially tune the prediction model based on performance \eqref{background:eq:rl:performance} rather than on system identification, implying that these learning parameters might grow to different values compared to the ones belonging to the true underlying dynamics. Finally, also constraints $h_{\bm\theta}$ \eqref{methodology:eq:mpc-func-approx:h} get parameterised, since $\tilde{\rho}_\text{crit}$ appears in it; see \eqref{background:eq:mpc-rm:ineq-constraints:h} and \eqref{background:eq:equivalent-flow-constraints}.
\begin{remark}
    In general, a parametrisation that makes use of quadratic terms with fixed setpoints, as in \eqref{methodology:eq:mpc-func-approx:stage-cost} and
    \eqref{methodology:eq:mpc-func-approx:terminal-cost}, can lead to suboptimal MPC policies in low-demand, free-flow regimes because it penalises low densities and low speeds as much as it penalises high densities and high speeds. At the same time, the theory at the foundation of MPC-based RL assumes the controller's parametrisation to be rich enough to adequately capture the real value function \cite{gros_2020_datadriven}. For this reason, in this work, we opt for the aforementioned adjustable quadratic cost terms. That being said, one could envision more involved and more performing learnable cost terms, especially in more complex and difficult traffic settings. For instance, asymmetric penalty functions are a valid choice, e.g., a left-sided Huber loss that quadratically penalises high traffic quantities, but penalises low ones only linearly.
\end{remark}
\begin{remark}
    The MPC parameters $\theta_\text{T},\theta_\text{V},\bm\Theta_\text{C}$ are disjoint from the RL stage cost weights $c_\text{T},c_\text{V},c_\text{C}$ of \eqref{methodology:eq:rl-stage-cost}. While they encode similar notions, there is in general no reason to expect the values of the former to converge to the values of the latter during learning. Rather, they will converge to those values that maximise the controller's closed-loop performance.
\end{remark}

The whole parametrisation is
\begin{equation}
    \bm\theta = \begin{bmatrix}
        \tilde{\rho}_\text{crit} &
        \tilde{a} &
        \theta_{\{\text{T},\text{V}\}} &
        \bm\Theta_\text{C}^\top &
        \bm\theta^{\{\rho,v,w\} \, \top}_{\{\lambda,\ell,\Gamma\}}
    \end{bmatrix}^\top.
\end{equation}
\Cref{methodology:table:mpc-func-approx:parametrisation} shows a summary of the parametrisation, reporting the scope and the space of each term. As reported in \Cref{numerical:table:parametrisation:init-vals} in the next section, these real spaces are often bounded in order to avoid undesired consequences, e.g., to preserve convexity of the parametric cost terms, or to prevent parameters that have some physical meaning from taking unrealistic values. 
\begin{table}
    \begin{center}
        \caption{Parametrisation $\bm\theta$ of the MPC function approximation \eqref{methodology:eq:mpc-func-approx:scheme}}
        \label{methodology:table:mpc-func-approx:parametrisation}
        \begin{tabular}{ccc}
            \hline Symbol & Scope & Space \\ \hline
            $\tilde{\rho}_\text{crit}$ & model, cost, constraint & $\mathbb{R}$ \\
            $\tilde{a}$ & model & $\mathbb{R}$ \\
            $\theta_\text{T}$ & cost - TTS weight & $\mathbb{R}$ \\
            $\theta_\text{V}$ & cost - control variability weight & $\mathbb{R}$ \\
            $\bm\Theta_\text{C}$ & cost - slack weights & $\mathbb{R}^{\card*{\mathcal{O}}\left(N_\text{p} + 1\right)}$ \\
            $\bm\theta^{\{\rho,v\}}_{\{\lambda,\ell,\Gamma\}}$ & \{init., stage, term.\} cost - \{$\rho$, $v$\} weights & $\mathbb{R}^{\card*{\mathcal{S}}}$ \\
            $\bm\theta^w_{\{\lambda,\ell,\Gamma\}}$ & \{init., stage, term.\} cost - $w$ weights & $\mathbb{R}^{\card*{\mathcal{O}}}$ \\
            \hline
        \end{tabular}
    \end{center}
\end{table}

Scheme \eqref{methodology:eq:mpc-func-approx:scheme} yields the approximation $V_{\bm\theta} : \mathbb{R}^{n_x} \rightarrow \mathbb{R}$ of the value function as
\begin{mini}
    { \scriptstyle \hat{\bm{X}}_k, \hat{\bm{U}}_k, \bm\Sigma }{
        \eqref{methodology:eq:mpc-func-approx:obj}
    }{ \label{methodology:eq:mpc-func-approx:V} }{ V_{\bm\theta}(\bm{x}_k) = }
    \addConstraint{ \eqref{methodology:eq:mpc-func-approx:init-state}-\eqref{methodology:eq:mpc-func-approx:sigma-constraint}. }{  }{ }
\end{mini}
The value function \eqref{methodology:eq:mpc-func-approx:V} satisfies the fundamental equalities of the Bellman equations \cite{gros_2020_datadriven}, so that
\begin{mini}
    { \scriptstyle \hat{\bm{X}}_k, \hat{\bm{U}}_k, \bm\Sigma }{
        \eqref{methodology:eq:mpc-func-approx:obj}
    }{ \label{methodology:eq:mpc-func-approx:Q} }{ Q_{\bm\theta}(\bm{x}_k, \bm{u}_k) = }
    \addConstraint{ \eqref{methodology:eq:mpc-func-approx:init-state}-\eqref{methodology:eq:mpc-func-approx:sigma-constraint}, }{  }{ }
    \addConstraint{ \hat{\bm{u}}_{0|k} = \bm{u}_k. }{}{}
\end{mini}
\begin{argmini}
    { \scriptstyle \bm{u} }{ Q_{\bm\theta}(\bm{x}_k, \bm{u}) }{ \label{methodology:eq:mpc-func-approx:pi} }{ \pi_{\bm\theta}(\bm{x}_k) = }.
\end{argmini}
In practice, policy \eqref{methodology:eq:mpc-func-approx:pi} is found as the first optimal action $\hat{\bm{u}}_{0|k}^\star$ from the $\arg \min$ of \eqref{methodology:eq:mpc-func-approx:V}. The goal of this parametrisation is to give the MPC scheme \eqref{methodology:eq:mpc-func-approx:scheme} enough degrees of freedom to sufficiently adapt to the RL task at hand. The cardinal point is that $\bm\theta$ must be rich enough to allow the scheme to capture the optimal policy $\pi^\star$ \cite{gros_2020_datadriven}. Of course, the choice of parametrisation is not unique, and other solutions may be viable: similarly to NNs, where the topology of the network and the number of neurons and activation functions to used in each layer are arbitrary, here also different choices are possible. In the context of RM, the inclusion of some parameters in $\bm\theta$ comes from the knowledge of the system and its control, e.g., the system's high sensitivity to the dynamics parameters $a$, $\rho_\text{crit}$ and the cost weights $\theta_\text{T}$, $\theta_\text{V}$, $\bm\theta_\text{C}$. Other elements of the parametrisation, such as the additional quadratic terms, are instead dictated by the framework \cite{gros_2020_datadriven} to ensure a rich approximation scheme. However, unlike NNs, the benefit of this model-based approach is that knowledge about the system can be incorporated in the parametrisation rather trivially. Further discussion on different parametrisations and their impact on the learning outcome can be found in the appendix.

Beyond a proper selection of $\bm\theta$, other factors influence the suboptimality of the MPC-based RL solution. During training, the RL algorithm can only process a finite amount of data, whereas theoretical optimality is achieved only as this amount approaches infinity \cite{sutton_2018_reinforcement}. The MPC approximation scheme is also inherently suboptimal due to its nonlinearity, converging to global optimality only in the asymptotic limit, though the issue can be alleviated via, e.g., multi-start \cite{rawlings_2017_model}.

\subsection{Second-Order LSTD Q-learning} \label{sections:sub:2nd-lstd-q-learning}

In order to adjust the parametrisation $\bm\theta$ of \eqref{methodology:eq:mpc-func-approx:scheme}, a second-order least-squares temporal difference (LSTD) Q-learning algorithm \cite{lagoudakis_2002_leastsquares} is employed to find the policy $\pi_{\bm\theta}$ that minimises the closed-loop performance. Q-learning is one of the most well-known indirect, temporal difference algorithms available in RL. In essence, it searches for the parametrisation that best fits the action-value function $Q_{\bm\theta}$ to the observed data, with the aim of recovering via this approximation the unknown optimal $Q_\star$ and, indirectly from that, the optimal policy $\pi_\star$. It has also shown very promising results in the context of MPC \cite{esfahani_2021_approximate}. The second-order Newton's method, coupled with an experience replay buffer of the past observed transitions \cite{lin_1992_selfimproving} and a re-formulation of the Q-fitting problem as a least-squares, ensures faster convergence and higher sample efficiency compared to traditional first-order methods. For more details, see \cite{esfahani_2021_approximate}.

More concretely, given the approximation $Q_{\bm\theta}$ (irrespective of its nature, e.g., MPC, NN, etc.), Q-learning solves the least-squares Bellman residual problem \eqref{background:eq:rl:bellman-residual-problem} via the second-order gradient update 
\begin{equation} \label{methodology:eq:vanilla-q-learning-update}
    \bm\theta \leftarrow \bm\theta - \alpha  \bm{H}^{-1} \bm{p},
\end{equation}
where $\alpha \ge 0$ is the learning rate, and the gradient $\bm{p}$ and Hessian $\bm{H}$ are found as
\begin{align}
    \bm{p} ={} & - \sum_{i=1}^{m}{ \delta_i \nabla_{\bm\theta} Q_{\bm\theta}(\bm{s}_i, \bm{a}_i) },
    \label{methodology:eq:rl:gradient} \\
    \bm{H} ={} & \sum_{i=1}^{m}{\nabla_{\bm\theta} Q_{\bm\theta}(\bm{s}_i, \bm{a}_i) \nabla_{\bm\theta} Q_{\bm\theta}^\top(\bm{s}_i, \bm{a}_i) - \delta_i \nabla^2_{\bm\theta} Q_{\bm\theta}(\bm{s}_i, \bm{a}_i) },
    \label{methodology:eq:rl:hessian}
\end{align}
with $m$ denoting the experience sample batch size and $\delta_i$ the TD error \eqref{background:eq:rl:td-error}. The update rule \eqref{methodology:eq:vanilla-q-learning-update} requires differentiation of the MPC scheme \eqref{methodology:eq:mpc-func-approx:Q} with respect to $\bm\theta$ to find $\nabla_{\bm\theta} Q_{\bm\theta}$ and $\nabla^2_{\bm\theta} Q_{\bm\theta}$. A nonlinear programming sensitivity analysis demonstrates how these quantities can be computed through differentiation of the Lagrangian of \eqref{methodology:eq:mpc-func-approx:Q} \cite{buskens_2001_sensitivity}. Prior to any update, the computed Hessian matrix is modified to be positive-definite by the addition of a multiple of the identity matrix \cite{nocedal_2006_numerical}. Furthermore, to impose lower and upper bounds on the parameters (as reported in Table \ref{numerical:table:parametrisation:init-vals}), \eqref{methodology:eq:vanilla-q-learning-update} is cast as the following optimisation problem
\begin{argmini}
    { \scriptstyle \Delta\bm\theta }{ \frac{1}{2} \Delta\bm\theta^\top \bm{H} \Delta\bm\theta + \alpha \bm{p}^\top \Delta\bm\theta}{ \label{methodology:eq:qp-update} }{ \Delta\bm\theta^\star = }
    \addConstraint{ \bm\theta_\text{lb} \le \bm\theta + \Delta\bm\theta \le \bm\theta_\text{ub}, }{}{}
    \addConstraint{ \Delta\bm\theta_\text{lb} \le \Delta\bm\theta \le \Delta\bm\theta_\text{ub}. }{}{}
\end{argmini}
This formulation allows to limit the rate of change of each parameter with $\Delta\bm\theta_\text{lb}$ and $\Delta\bm\theta_\text{ub}$. The parametrisation is then updated as $\bm\theta \leftarrow \bm\theta + \Delta\bm\theta^\star$. Note that, compared to \eqref{methodology:eq:mpc-func-approx:scheme}, \eqref{methodology:eq:qp-update} adds negligible computational complexity as it is a convex quadratic problem and is usually solved at a lower, batch-update frequency (as explained in \Cref{sections:sub:sub:rl-numerical}).

Lastly, as in \cite{zanon_2021_safe}, exploratory behaviour is ensured during learning by adding to the MPC objective \eqref{methodology:eq:mpc-func-approx:obj} the perturbation term $\bm{q}^\top \bm{u}_0$, with $\bm{q}$ randomly chosen from, e.g., a normal distribution. When properly calibrated, this perturbation on the first action ensures that enough exploration is added on top of the policy in order to prevent the Q-learning agent from getting stuck in very suboptimal local minima.

\section{Numerical Case Study}  \label{sections:numerical}

To examine the performance of the proposed method for RM control, we implement and simulate the algorithm when applied to a highway network benchmark.

\subsection{Configuration}

\subsubsection{Network Environment}

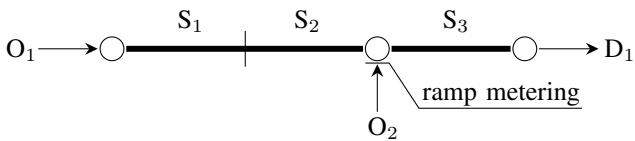
\begin{figure}
    \centering
    \begin{tikzpicture}[x=0.45pt,y=0.45pt,yscale=-1,xscale=1]
    \draw (30,40) -- (77,40);
    \draw [shift={(80,40)}, rotate = 180, fill={rgb,255:red,0;green,0;blue,0}, line width=0.08, draw opacity=0] (10.72,-5.15) -- (0,0) -- (10.72,5.15) -- (7.12,0) -- cycle;
    \draw [line width=2.25] (104,40) -- (303,40);
    \draw [line width=2.25] (327,40) -- (427,40);
    \draw (451,40) -- (498,40);
    \draw [shift={(501,40)}, rotate = 180, fill={rgb,255:red,0;green,0;blue,0}, line width=0.08, draw opacity=0] (10.72,-5.15) -- (0,0) -- (10.72,5.15) -- (7.12,0) -- cycle;
    \draw (204,25) -- (204,55);
    \draw (315,93) -- (315,56);
    \draw [shift={(315,53)}, rotate = 90, fill={rgb,255:red,0;green,0;blue,0}, line width=0.08,draw opacity=0] (10.72,-5.15) -- (0,0) -- (10.72,5.15) -- (7.12,0) -- cycle;
    \draw (305,52) -- (325,52) -- (350,90) -- (487,90);

    \draw   (82,40) .. controls (82,34.48) and (86.48,30) .. (92,30) .. controls (97.52,30) and (102,34.48) .. (102,40) .. controls (102,45.52) and (97.52,50) .. (92,50) .. controls (86.48,50) and (82,45.52) .. (82,40) -- cycle ;
    \draw   (305,40) .. controls (305,34.48) and (309.48,30) .. (315,30) .. controls (320.52,30) and (325,34.48) .. (325,40) .. controls (325,45.52) and (320.52,50) .. (315,50) .. controls (309.48,50) and (305,45.52) .. (305,40) -- cycle ;
    \draw   (429,40) .. controls (429,34.48) and (433.48,30) .. (439,30) .. controls (444.52,30) and (449,34.48) .. (449,40) .. controls (449,45.52) and (444.52,50) .. (439,50) .. controls (433.48,50) and (429,45.52) .. (429,40) -- cycle ;

    \draw (1,30) node [anchor=north west, inner sep=0.75pt, align=left] {$\text{O}_{1}$};
    \draw (504,30) node [anchor=north west, inner sep=0.75pt, align=left] {$\text{D}_{1}$};
    \draw (145,5) node [anchor=north west, inner sep=0.75pt, align=left] {$\text{S}_{1}$};
    \draw (243,5) node [anchor=north west, inner sep=0.75pt, align=left] {$\text{S}_{2}$};
    \draw (368,5) node [anchor=north west, inner sep=0.75pt, align=left] {$\text{S}_{3}$};
    \draw (305,95) node [anchor=north west, inner sep=0.75pt, align=left] {$\text{O}_{2}$};
    \draw (351,66) node [anchor=north west, inner sep=0.75pt, align=left] {ramp metering};
\end{tikzpicture}
    \caption{Structure of the three-segment highway network}
    \label{numerical:fig:traffic-network}
\end{figure}

Let us consider a simple highway traffic network taken from \cite{hegyi_2004_model} and pictured in \cref{numerical:fig:traffic-network}, which will serve as the RL environment for the task at hand. The network consists of three segments, each 1 km in length and with two lanes. The first segment $\text{S}_1$ is supplied by the uncontrolled mainstream origin $\text{O}_1$, which simulates incoming traffic from the unmodelled upstream region of the network, and is characterized by its capacity $C_1$. Between segments $\text{S}_2$ and $\text{S}_3$, additional traffic can enter the network via the on-ramp $\text{O}_2$. This origin has capacity $C_2$ and offers the only control measure in order to avoid congestion in the network. However, we would also like to keep the queue length on this on-ramp to 50 vehicles or fewer. Lastly, traffic exits the network via the only available destination $\text{D}_1$ with congested outflow.

\begin{figure}
    \centering
    \input{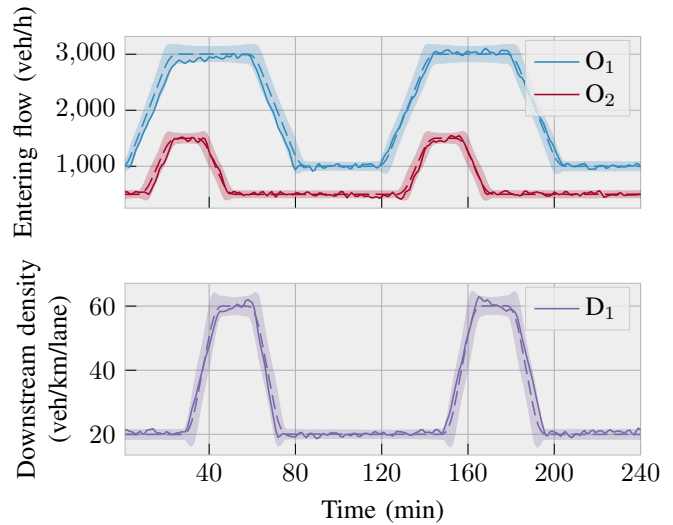}
    \caption{External inputs affecting the network's dynamics, in the form of the demands at the origins (upper) and the congestion scenario at the destination (lower). The shaded areas represent the 2-standard deviation ranges from which the random demands are sampled, whereas the solid lines represent one random sample for each.}
    \label{numerical:fig:demands}
\end{figure}

As described in \Cref{sections:sub:metanet}, the network environment is subject to one external input per origin and destination. \Cref{numerical:fig:demands} depicts the demands at both origins, starting low and then increasing to near capacity within 20 minutes. With some delay, also congestion at the destination rapidly appears. These profiles are intended to simulate a peak-hour-like traffic (e.g., morning or evening rush), and are devised in such a way to induce a variable degree of congestion in the network, if the on-ramp is not properly controlled. At the start of each training episode, these two-peak profiles are generated randomly to introduce a degree of variability in the task, aimed at enhancing the agent's generalisation to a wider variety of congestion scenarios. Nevertheless, generalisation in RL is challenging \cite{ghosh_2021_why_generalization}, and a performance drop is anticipated were the controller to face a new unforeseen scenario, e.g., with a different number of peaks. At the same time, our approach exhibits an inherent level of robustness: it incorporates an online learning process that can seamlessly integrate novel training data on the fly, and at its core is a model-based controller, which generally extrapolates better to new situations in comparison to its model-free counterparts.

\subsubsection{Model Predictive Control}

The controller \eqref{methodology:eq:mpc-func-approx:scheme} is deployed to control the on-ramp flow with the aim of avoiding congestion in the traffic network environment. In line with other similar benchmarks in the literature, e.g., \cite{hegyi_2002_optimal,hegyi_2004_model,hegyi_2005_model}, $M = 6$, meaning that, while the process dynamics are stepped every 10 seconds, the controller is run only once per minute. The prediction horizon is set to $N_\text{p} = 24$ (which is in the order of the average travel time through the network), and the control horizon to $N_\text{c}$ = 3.

To showcase the ability of RL to automatically improve the performance of the MPC controller, the parametrisation $\bm\theta$ of the controller is poorly initialised on purpose. In particular, in order to replicate the effects of a poor system identification phase, the parameters of the prediction model are initialised with a 30\% error with respect to their true values, i.e., $\tilde{\rho}_\text{crit} = 0.7 \rho_\text{crit}$, $\tilde{a} = 1.3 a$ (both learnable), and  $1.3 v_\text{free}$ (fixed). Furthermore, the remaining parametrisation of the objective function of the MPC optimisation problem is left untuned, imitating a suboptimal, low-expertise design process of the controller. \Cref{numerical:table:parametrisation:init-vals} reports the initial values of each learnable parameter. The other non-learnable parameters in \eqref{methodology:eq:mpc-func-approx:scheme} are set to the same values as in the real process. The queue threshold on $O_2$ is set to $w_\text{max} = \qty{50}{\vehicle}$. Lastly, the setpoints are set to $\rho_\text{sp} = 0.7 \rho_\text{crit}$, $v_\text{sp} = 1.3 v_\text{free}$, and the normalization coefficients to $\rho_\text{max} = \qty{180}{\unit{\vehicle\per\kilo\metre\per\lane}}$, $w_\text{max} = \qty{50}{\vehicle}$, and $v_\text{max} = 1.3 v_\text{free}$.
\begin{table}
    \begin{center}
        \caption{Initial values, bounds and dimensions of the parametrisation $\bm\theta$ of \eqref{methodology:eq:mpc-func-approx:scheme}}
        \label{numerical:table:parametrisation:init-vals}
        \begin{tabular}{cccc}
            \hline Symbol                                         & Initial value & Bounds             & Dimension                               \\ \hline
            $\tilde{\rho}_\text{crit}$                            & 23.45         & $[10,162]$         & \unit{\vehicle\per\kilo\metre\per\lane} \\
            $\tilde{a}$                                           & 2.4271        & $[1.1,3]$          & -                                       \\
            $\theta_\text{T}$                                     & 1             & $[10^{-3},\infty)$ & \unit{\per\vehicle\per\hour}            \\
            $\theta_\text{V}$                                     & 160000        & $[10^{-3},\infty)$ & \unit{\square\vehicle\per\square\hour}  \\
            $\bm\Theta_\text{C}$                                  & 5             & $[10^{-3},\infty)$ & \unit{\per\vehicle}                     \\
            $\bm\theta^{\{\rho, v, w\}}_{\lambda}$                & 1             & $(-\infty,\infty)$ & -                                       \\
            $\bm\theta^{\{\rho, v, w\}}_{\{\ell, \Gamma\}}$       & 1             & $[10^{-6},\infty)$ & -                                       \\
            \hline
        \end{tabular}
    \end{center}
\end{table}

\subsubsection{Reinforcement Learning} \label{sections:sub:sub:rl-numerical}

The agent is trained in an episodic fashion, with each episode lasting $T_\text{fin}=\qty{4}{\hour}$, i.e., each episode features two peaks in demands and congestion in a row (for reference, see \cref{numerical:fig:demands}). At the end of each episode, the demands are generated anew randomly, and the state of the network is reset to steady-state in order to avoid unreasonable accumulation of vehicles in queues from past episodes. With the same frequency, the parametrisation $\bm\theta$ is updated, and a new episode is started, till a termination condition of the learning process is met, typically in the form of a maximum number of iterations. Here, we train our agent for 80 episodes. With the aforementioned update frequency, the proposed learning setup is off-policy, since $\theta$ remains fixed throughout each episode, ensuring stable data collection.

The stage cost function $L(\bm{x}_k, \bm{u}_k)$, with which the traffic network feeds a cost signal back to the agent according to \eqref{methodology:eq:rl-stage-cost}, has coefficients $c_\text{T} = 5$, $c_\text{V} = 1600$, and $c_\text{C} = 5$.

Based on results obtained from initial simulations, the algorithm's hyperparameters are set as follows. The discount factor is set to $\gamma = 0.98$. The learning rate is initially set to $\alpha = 0.925$, and decayed by a multiplicative factor of 0.925 at the end of each update. The maximum update change of each parameter is set to 30\% of its current value, i.e., $\Delta\bm\theta_\text{ub} = -\Delta\bm\theta_\text{lb} = 0.3 \bm\theta$. As aforementioned, a replay buffer is used to store past observations in memory and re-use them. In particular, the buffer size is set up to store transitions from the 10 past episodes and, before performing an update, a batched sample half its size is drawn from this buffer. In turn, half of this batch is dedicated to containing information from the latest two and a half episodes, whereas the remaining half is sampled uniformly at random from even older transitions. All together, these measures contribute to stabilising the learning process.

Lastly, exploration is induced as described in \Cref{sections:sub:2nd-lstd-q-learning} by sampling the cost perturbation from a normal distribution in an $\varepsilon$-greedy fashion, i.e., $\bm{q} \sim \mathcal{N}(0, \sigma_q)$ with probability $\varepsilon$; otherwise, $\bm{q} = 0$. The exploration strength is $\sigma_q = 0.025$, and the exploration probability $\varepsilon = 0.5$. Both are decayed by half at the end of each episode.

\Cref{numerical:table:hyperparameters} summarises all the hyper-parameters and non-learnable parameters of the traffic environment and the MPC controller, as found in \cite{hegyi_2004_model}, as well as of the RL algorithm.
\begin{table*}
    \begin{center}
        \caption{Hyper- and other non-learnable parameters}
        \label{numerical:table:hyperparameters}
        \begin{tabular}{cc|cccccccccc}
            \hline
            \multirow{3}{*}{METANET} & Symbol    & $T$            & $\tau$         & $\eta$                             & $\kappa$                                   & $\mu$                       & $\rho_\text{max}$                       & $\rho_\text{crit}$                      & $v_\text{free}$                              & $a$               & $C_{\{1,2\}}$            \\
                                     & Value     & 10             & 18             & 60                                 & 40                                         & 0.0122                      & 180                                     & 33.5                                    & 102                                          & 1.867             & \{3500, 2000\}           \\
                                     & Dimension & \unit{\second} & \unit{\second} & \unit{\square\kilo\metre\per\lane} & \unit{\vehicle\per\kilo\metre\per\lane}    & -                           & \unit{\vehicle\per\kilo\metre\per\lane} & \unit{\vehicle\per\kilo\metre\per\lane} & \unit{\kilo\metre\per\hour}                  & -                 & \unit{\vehicle\per\hour} \\ \hline
            \multirow{3}{*}{MPC}     & Symbol    & $M$            & $N_\text{p}$   & $N_\text{c}$                       & $w_\text{max}$                             & $v_\text{max}$              & $\rho_\text{sp}$                        & $v_\text{sp}$                           &                                              &                   &                          \\
                                     & Value     & 6              & 24             & 3                                  & 50                                         & 132.6                       & 23.45                                   & 132.6                                   &                                              &                   &                          \\
                                     & Dimension & -              & -              & -                                  & \unit{\vehicle}                            & \unit{\kilo\metre\per\hour} & \unit{\vehicle\per\kilo\metre\per\lane} & \unit{\kilo\metre\per\hour}             &                                              &                   &                          \\ \hline
            \multirow{2}{*}{RL}      & Symbol    & $c_\text{T}$   & $c_\text{V}$   & $c_\text{C}$                       & batch size                                 & $\gamma$                    & $\alpha$                                & $\alpha$ decay                          & $\Delta\bm\theta_{\{\text{lb},\text{ub}\}}$  & $\sigma_{\bm{q}}$ & $\varepsilon$            \\
                                     & Value     & 5              & 1600           & 5                                  & 5 episodes                                 & 0.98                        & 0.925                                   & 0.925                                   & \{-0.3$\theta$,0.3$\theta$\}                                & 0.025             & 0.5                      \\ \hline
        \end{tabular}
    \end{center}
\end{table*}

\subsubsection{Setup}

The simulations were run on a Ubuntu 20.04.6 server equipped with 16 AMD EPYC 7252 (3.1 GHz) processors and 252GB of RAM, and implemented in Python 3.11.4. The nonlinear optimisation problems were formulated and solved with the symbolic framework CasADi \cite{andersson_2019_casadi} and its interface to the IPOPT solver \cite{wachter_2006_implementation}. The source code and simulation results are open and available in the following repository: \url{https://github.com/FilippoAiraldi/mpcrl-for-ramp-metering}.

\subsection{Comparison}

Alongside the proposed MPC-based RL controller, we implement and simulate on the same setup three other policies. These include another learning-based approach based on DDPG, a DRL algorithm that has been proven effective in freeway control \cite{wang_2022_integrated,sun_2024_novel}, and two other non-learning solutions: the classical non-learning MPC formulation \eqref{background:eq:mpc-rm:scheme} \cite{hegyi_2004_model}, as well as the well-known local ramp metering PI-ALINEA \cite{papageorgiou_1991_alinea,wang_2014_local}. To promote a fair comparison, all of the tested controllers suffer from the same inexact knowledge of the traffic parameters.

The relevant hyper-parameters for DDPG are taken from \cite{sun_2024_novel}. The PI-ALINEA controller is equipped with a queue management strategy to take into account the queue length constraint \cite{spiliopoulou_2010_queue}, and its gains are fine-tuned to this specific task via Bayesian Optimisation.

\subsection{Results}

We evaluate the performance of all the aforementioned methods deployed on the traffic network environment and average the results over 15 simulations with different seeds to iron out randomness due to, e.g., exploration. Figures that are shown and discussed next report the average results; however, 95\% confidence intervals are shown only for our proposed method to reduce visual clutter.

\begin{figure}
    \centering
    \input{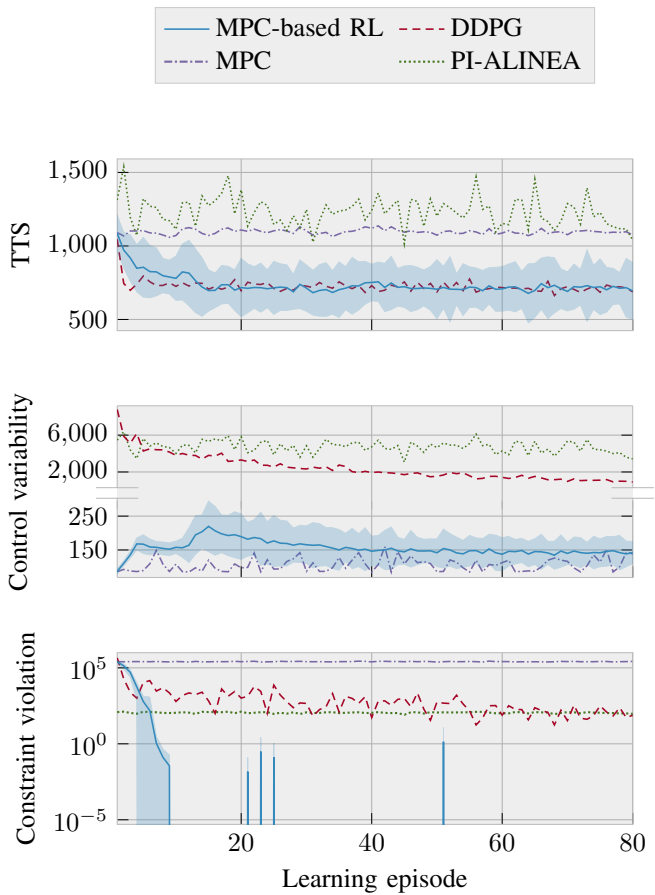}
    \caption{Evolution of the three contributions to the RL cost \eqref{methodology:eq:rl-stage-cost} during the learning process, namely, from top to bottom, the Total-Time-Spent (TTS), variability of the control action, and violation of the maximum queue constraint on $\text{O}_2$}
    \label{numerical:fig:learning-costs}
\end{figure}

\Cref{numerical:fig:learning-costs} shows, for each method, the total RL cost \eqref{methodology:eq:rl-stage-cost} incurred in each episode, i.e., $\sum_{k=1}^{T_\text{fin}/T}{ L(\bm{x}_k,\bm{u}_k) }$, subdivided in its three contributions, and how these evolve during learning. Due to the untuned objective weights and the 30\% mismatches in the predictive model's parameters, the initial MPC-based RL controller is poorly performing and cannot reliably avoid congestion and constraint violations, as can be seen from the costs in the first episodes. However, despite these initial shortcomings, it can be noticed that in the next 20 episodes, by exclusively leveraging the observed transitions via Q-learning, the controller is able to achieve a substantial improvement of the Total-Time-Spent cost, i.e., the time spent by vehicles in the network on average, from around \qtyrange{1100}{700}{\vehicle\hour} (or a $35\%$ reduction). At the same time, as constraint violations of $w_\text{max}$ in $\text{O}_2$ induce the most severe cost realization by several orders of magnitude, one can see that the agent is even quicker in learning to avoid such a constraint-violating behaviour within the first 5 episodes, despite the fact it has no knowledge of the exact traffic dynamics parameters.

\begin{figure}
    \centering
    \input{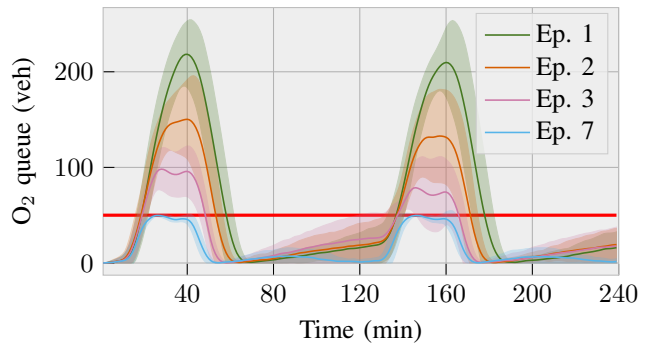}
    \caption{On average, the policy $\pi_{\bm\theta}$ gets better at avoiding constraint violations as it learns (the red line represents the threshold $w_\text{max}$ imposed on the queue on the on-ramp $\text{O}_2$)}
    \label{numerical:fig:learning-queue-violation}
\end{figure}

This phenomenon can be further appreciated in \cref{numerical:fig:learning-queue-violation}, which shows the progression of the average queue in the on-ramp $\text{O}_2$ with respect to its constraint, as the parametrised MPC scheme gets better and better at yielding a policy with less or no constraint violation. It must be noted here that, despite the non-stationary nature of the traffic environment, the agent is able to learn a policy that is robust to the variability of the randomly generated demand and congestion scenarios and, therefore, is capable of avoiding all constraint violations.

Further analysis of the results indicates that, while learning to satisfy the constraint on the on-ramp $\text{O}_2$ is directly beneficial to the reduction of cost related to violations, it also indirectly fosters better TTS, since longer queues in $\text{O}_2$ proportionally relate to a higher TTS. Nonetheless, improvements to the violation and TTS contributions seem to occur at the expense of the third contribution, i.e., variability of the control action, which clearly is on the rise during learning for the MPC-based RL agent. However, in the overall context of learning, one must notice that at convergence the sacrifice in control variability (of around 60 points) allows the agent to gain a larger improvement in TTS and to achieve zero constraint violations (which are indeed the largest contributors to the cost of the RL agent). Furthermore, as the learning proceeds past the $16^{\text{th}}$ episode and the other two costs have settled, the agent is also able to achieve further reduction of the cost associated to the control action variability.

The results from the other methods appear to corroborate the usefulness of data-driven adaptability in such a context. The non-learning MPC formulation is stuck at the initial poor performance of its RL-enhanced counterpart, with high TTS and constraint violations. On the contrary, thanks to its queue management strategy, PI-ALINEA is able to curb violations, but at the expense of a much higher control variability (since the strategy frequently requires to switch control action abruptly) and higher TTS. The DDPG agent follows a similar trend to our method in quickly learning to reduce TTS. However, variability and violations are decreasing at a substantially lower rate. This is not unexpected, as the NN approximation of this agent has significantly more learnable parameters (namely, $293\,380$ vs. $53$) and is thus less data-efficient and requires more episodes to establish convergence.

\begin{figure}
    \centering
    \input{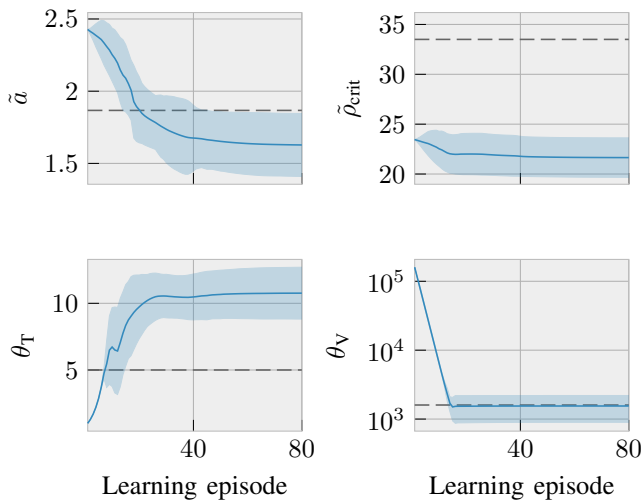}
    \caption{Evolution of a subset of $\bm\theta$ during the learning process (the dashed grey lines represent, in the top plots, the true values of the parameters $a$ and $\rho_\text{crit}$, and, in the bottom ones, the constants $c_\text{T}$ and $c_\text{V}$)}
    \label{numerical:fig:learning-parameters}
\end{figure}

As discussed in \Cref{sections:introduction}, an advantage of the proposed MPC-RL approach over DRL is that its learning process can be more readily inspected and explained via the evolution of its parametrisation $\bm\theta$, reported in \Cref{numerical:fig:learning-parameters}. In particular
\begin{itemize}
    \item the evolution of $\theta_\text{T}$ and $\theta_\text{V}$ agrees with the observations made in \cref{numerical:fig:learning-costs}, that is, the agent learns to favour the TTS cost over the control action variability, and thus increases the weight of the former at the expense of the latter
    \item the parameter $\tilde{a}$, which the METANET model is known to be very sensitive to, develops at first towards its true value $a$, but then convergences to a slightly lower value, resulting in a less pronounced, less aggressive equilibrium speed $V_\text{e}$ \eqref{background:eq:equilibrium-speed}, i.e., favouring predictions of lower speeds at lower densities, and higher speeds at higher densities
    \item $\tilde{\rho}_\text{crit}$ grows in the opposite direction of its true value $\rho_\text{crit}$, thus settling for a prediction model that is pessimistic towards congestion, i.e., it tends to overestimate congestion scenarios, as well as resulting in a tighter constraint $h_o^1$ \eqref{background:eq:equivalent-flow-constraints}, which in turn restricts the control action further.
\end{itemize}

\begin{figure}
    \centering
    \input{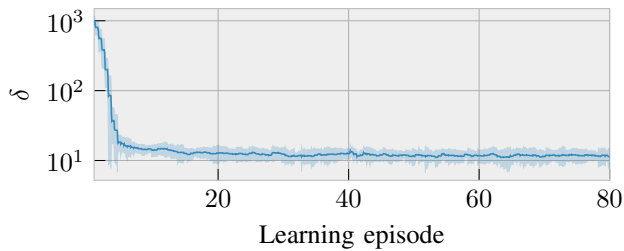}
    \caption{Moving average of the TD error during the learning process ($\text{window length} = 239$, i.e., number of transitions per episode)}
    \label{numerical:fig:td-error}
\end{figure}

\Cref{numerical:fig:td-error} depicts the TD error during the learning process, whose moving average converges to smaller and smaller values as the episodes increase. This is a good indication that the MPC scheme \eqref{methodology:eq:mpc-func-approx:scheme}, with its parametrisation $\bm\theta$, is able to provide a reliable approximation $Q_{\bm\theta}$ of the true unknown action-value function (and, indirectly, of the optimal policy). Yet, it is important to note that the convergence of the TD error, while consistently diminishing, does not reach an absolute zero. This observation can be attributed to factors such as imperfect parametrisation and the inherent stochasticity present in the environment, which fluctuates in its demands, and thus renders the prediction of the value functions more challenging. Besides these issues, another challenge related to the TD error is its sparsity. Specifically, the instantaneous TD signal remains low and largely uninformative for most of the episode, only to spike significantly during congestion events. This sparsity can lead to inefficient updates if not properly addressed. Therefore, we advocate that the use of experience replay and batch updates is essential to smooth the learning process and mitigate the risk of destabilization due to abrupt parameter adjustments.

\begin{figure*}
    \centering
    \begin{tikzpicture}

    \definecolor{darkgray178}{RGB}{178,178,178}
    \definecolor{silver188}{RGB}{188,188,188}
    \definecolor{whitesmoke238}{RGB}{238,238,238}

    \begin{groupplot}[
            group style={
                    group size=2 by 3, vertical sep=0.5cm, horizontal sep=0.5cm,
                },
            width=\axisdefaultwidth,
            height=0.852*\axisdefaultheight,
        ]
        \nextgroupplot[
        axis background/.style={fill=whitesmoke238},
        axis line style={silver188},
        scaled x ticks=manual:{}{\pgfmathparse{#1}},
        tick pos=left,
        tick scale binop=\times,
        title={Episode 1},
        x grid style={darkgray178},
        xmajorgrids,
        xmin=1, xmax=240,
        xtick style={color=black},
        xticklabels={},
        y grid style={darkgray178},
        ymajorgrids,
        ymin=0.5, ymax=3.5,
        ytick style={color=black},
        ytick={1,2,3},
        yticklabels={\(\displaystyle \text{S}_1\),\(\displaystyle \text{S}_2\),\(\displaystyle \text{S}_3\)},
        ]
        \addplot graphics [includegraphics cmd=\pgfimage,xmin=1, xmax=240, ymin=0.5, ymax=3.5] {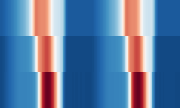};

        \nextgroupplot[
        axis background/.style={fill=whitesmoke238},
        axis line style={silver188},
        colorbar,
        colorbar style={
            ylabel={$\rho$ (veh/km/lane)},
            ytick={0,50,100},
        },
        colormap={mymap}{[1pt]
                rgb(0pt)=(0.0196078431372549,0.188235294117647,0.380392156862745);
                rgb(1pt)=(0.129411764705882,0.4,0.674509803921569);
                rgb(2pt)=(0.262745098039216,0.576470588235294,0.764705882352941);
                rgb(3pt)=(0.572549019607843,0.772549019607843,0.870588235294118);
                rgb(4pt)=(0.819607843137255,0.898039215686275,0.941176470588235);
                rgb(5pt)=(0.968627450980392,0.968627450980392,0.968627450980392);
                rgb(6pt)=(0.992156862745098,0.858823529411765,0.780392156862745);
                rgb(7pt)=(0.956862745098039,0.647058823529412,0.509803921568627);
                rgb(8pt)=(0.83921568627451,0.376470588235294,0.301960784313725);
                rgb(9pt)=(0.698039215686274,0.0941176470588235,0.168627450980392);
                rgb(10pt)=(0.403921568627451,0,0.12156862745098)
            },
        point meta max=105.908965898626,
        point meta min=0,
        scaled x ticks=manual:{}{\pgfmathparse{#1}},
        scaled y ticks=manual:{}{\pgfmathparse{#1}},
        tick pos=left,
        tick scale binop=\times,
        title={Episode 80},
        x grid style={darkgray178},
        xmajorgrids,
        xmin=1, xmax=240,
        xtick style={color=black},
        xtick={40,80,120,160,200,240},
        xticklabel style={text opacity=0},
        y grid style={darkgray178},
        ymajorgrids,
        ymin=0.5, ymax=3.5,
        ytick style={color=black},
        yticklabels={}
        ]
        \addplot graphics [includegraphics cmd=\pgfimage,xmin=1, xmax=240, ymin=0.5, ymax=3.5] {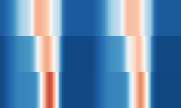};

        \nextgroupplot[
        axis background/.style={fill=whitesmoke238},
        axis line style={silver188},
        scaled x ticks=manual:{}{\pgfmathparse{#1}},
        tick pos=left,
        tick scale binop=\times,
        x grid style={darkgray178},
        xmajorgrids,
        xmin=1, xmax=240,
        xtick style={color=black},
        xticklabels={},
        y grid style={darkgray178},
        ymajorgrids,
        ymin=0.5, ymax=3.5,
        ytick style={color=black},
        ytick={1,2,3},
        yticklabels={\(\displaystyle \text{S}_1\),\(\displaystyle \text{S}_2\),\(\displaystyle \text{S}_3\)},
        ]
        \addplot graphics [includegraphics cmd=\pgfimage,xmin=1, xmax=240, ymin=0.5, ymax=3.5] {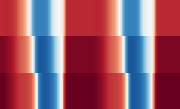};

        \nextgroupplot[
        axis background/.style={fill=whitesmoke238},
        axis line style={silver188},
        colorbar,
        colorbar style={
            ylabel={$v$ (km/h)},
            ytick={0,50,100},
        },
        colormap={mymap}{[1pt]
                rgb(0pt)=(0.0196078431372549,0.188235294117647,0.380392156862745);
                rgb(1pt)=(0.129411764705882,0.4,0.674509803921569);
                rgb(2pt)=(0.262745098039216,0.576470588235294,0.764705882352941);
                rgb(3pt)=(0.572549019607843,0.772549019607843,0.870588235294118);
                rgb(4pt)=(0.819607843137255,0.898039215686275,0.941176470588235);
                rgb(5pt)=(0.968627450980392,0.968627450980392,0.968627450980392);
                rgb(6pt)=(0.992156862745098,0.858823529411765,0.780392156862745);
                rgb(7pt)=(0.956862745098039,0.647058823529412,0.509803921568627);
                rgb(8pt)=(0.83921568627451,0.376470588235294,0.301960784313725);
                rgb(9pt)=(0.698039215686274,0.0941176470588235,0.168627450980392);
                rgb(10pt)=(0.403921568627451,0,0.12156862745098)
            },
        point meta max=100.320470851389,
        point meta min=0,
        scaled x ticks=manual:{}{\pgfmathparse{#1}},
        scaled y ticks=manual:{}{\pgfmathparse{#1}},
        tick pos=left,
        tick scale binop=\times,
        x grid style={darkgray178},
        xmajorgrids,
        xmin=1, xmax=240,
        xtick style={color=black},
        xtick={40,80,120,160,200,240},
        xticklabel style={text opacity=0},
        y grid style={darkgray178},
        ymajorgrids,
        ymin=0.5, ymax=3.5,
        ytick style={color=black},
        yticklabels={}
        ]
        \addplot graphics [includegraphics cmd=\pgfimage,xmin=1, xmax=240, ymin=0.5, ymax=3.5] {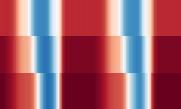};

        \nextgroupplot[
            axis background/.style={fill=whitesmoke238},
            axis line style={silver188},
            tick pos=left,
            tick scale binop=\times,
            x grid style={darkgray178},
            xlabel={Time (min)},
            xmajorgrids,
            xmin=1, xmax=240,
            xtick={40,80,120,160,200,240},
            xtick style={color=black},
            y grid style={darkgray178},
            ymajorgrids,
            ymin=0.5, ymax=3.5,
            ytick style={color=black},
            ytick={1,2,3},
            yticklabels={\(\displaystyle \text{S}_1\),\(\displaystyle \text{S}_2\),\(\displaystyle \text{S}_3\)},
        ]
        \addplot graphics [includegraphics cmd=\pgfimage,xmin=1, xmax=240, ymin=0.5, ymax=3.5] {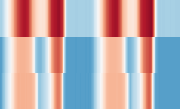};

        \nextgroupplot[
        axis background/.style={fill=whitesmoke238},
        axis line style={silver188},
        colorbar,
        colorbar style={ylabel={
            $q$ (veh/h)},
            ytick={0,2000,4000},
        },
        colormap={mymap}{[1pt]
                rgb(0pt)=(0.0196078431372549,0.188235294117647,0.380392156862745);
                rgb(1pt)=(0.129411764705882,0.4,0.674509803921569);
                rgb(2pt)=(0.262745098039216,0.576470588235294,0.764705882352941);
                rgb(3pt)=(0.572549019607843,0.772549019607843,0.870588235294118);
                rgb(4pt)=(0.819607843137255,0.898039215686275,0.941176470588235);
                rgb(5pt)=(0.968627450980392,0.968627450980392,0.968627450980392);
                rgb(6pt)=(0.992156862745098,0.858823529411765,0.780392156862745);
                rgb(7pt)=(0.956862745098039,0.647058823529412,0.509803921568627);
                rgb(8pt)=(0.83921568627451,0.376470588235294,0.301960784313725);
                rgb(9pt)=(0.698039215686274,0.0941176470588235,0.168627450980392);
                rgb(10pt)=(0.403921568627451,0,0.12156862745098)
            },
        point meta max=4444.17647731368,
        point meta min=0,
        tick pos=left,
        tick scale binop=\times,
        x grid style={darkgray178},
        xlabel={Time (min)},
        xmajorgrids,
        xmin=1, xmax=240,
        xtick={40,80,120,160,200,240},
        xtick style={color=black},
        y grid style={darkgray178},
        ymajorgrids,
        ymin=0.5, ymax=3.5,
        ytick style={color=black},
        scaled y ticks=manual:{}{\pgfmathparse{#1}},
        yticklabels={}
        ]
        \addplot graphics [includegraphics cmd=\pgfimage,xmin=1, xmax=240, ymin=0.5, ymax=3.5] {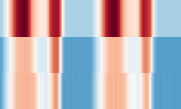};
    \end{groupplot}
\end{tikzpicture}
    \caption{Differences in traffic quantities of the network's three segments between the first episode and the last episode of the learning process, averaged across the 15 simulation runs}
    \label{numerical:fig:traffic-quantities}
\end{figure*}

Finally, \Cref{numerical:fig:traffic-quantities} reports the evolution of the traffic quantities of the three segments (i.e., density $\rho$, speed $v$, and flow $q$) that make up the network. Interestingly, it can be noticed how, compared to the initial episode, the final controller is able to better prevent the congestion in the last segment from propagating backwards through the network, and from causing speed drops in the first segment.

\section{Conclusions}  \label{sections:conclusions}

In this paper, we have proposed a novel learning-based and model-based approach to the ramp metering problem that combines MPC and RL. The two frameworks are integrated in such a way as to promote the strengths of each while countering the disadvantages by exploiting the parametrised MPC scheme as a function approximation of the action-value function and leveraging RL to adjust the parametrisation based on observed data to improve closed-loop performance. Even with wrong model parameters and a poorly tuned initial controller, the proposed methodology shows a remarkable ability in learning to improve traffic control performance and satisfy constraints in an automatic, data-driven fashion.

Future work will focus on
\begin{enumerate*}[label=\arabic*)]
    \item the validation of the proposed methodology with microscopic traffic model simulators,
    \item the study of formal guarantees on stability and recursive feasibility for the proposed approach during learning and at convergence,
    \item more complex parametrisations of the MPC scheme, e.g., with neural networks, in order to better address the nonlinear nature of the METANET framework and to better capture the shape of the true action-value function, as well as
    \item extending the current approach by integrating variable speed limits (VSLs) with ramp metering, a coordinated control strategy that has been shown to achieve better results than ramp metering alone, especially in high demand/density regimes.
\end{enumerate*}

\bibliographystyle{IEEEtran}
\bibliography{IEEEabrv,references_long}

\begin{IEEEbiography}[{\includegraphics[width=1in,height=1.25in,clip,keepaspectratio]{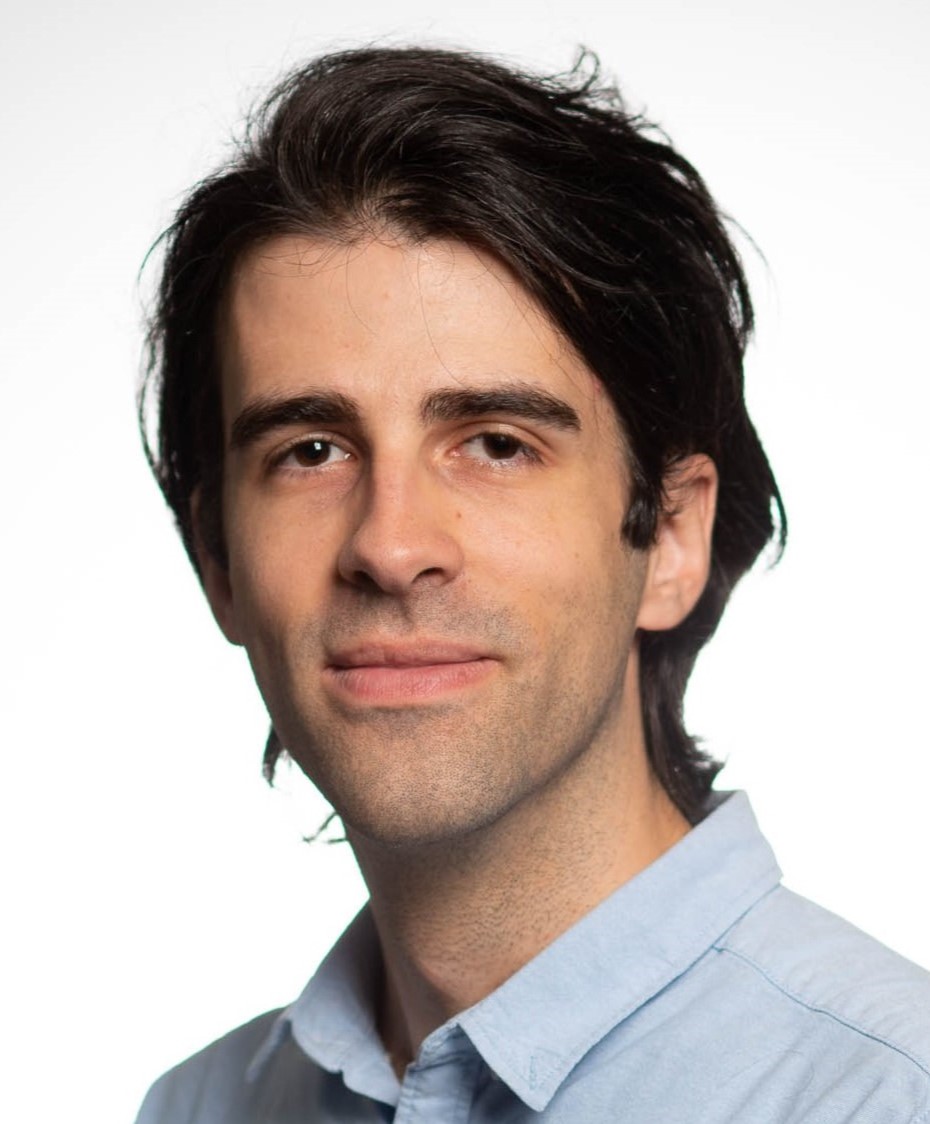}}]{Filippo Airaldi}
    received the B.Sc. and M.Sc. degrees from the Polytechnic University of Turin, Italy, in 2017 and 2019, respectively. He is currently a Ph.D. candidate at the Delft Center for Systems and Control, Delft University of Technology, The Netherlands.

    His research interests include model predictive control, reinforcement learning, and other machine learning techniques, and in particular in their combination with learning-based control.
\end{IEEEbiography}
\begin{IEEEbiography}[{\includegraphics[width=1in,height=1.25in,clip,keepaspectratio]{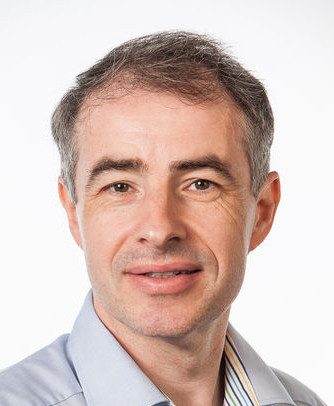}}]{Bart De Schutter}
    (Fellow, IEEE) received the Ph.D. degree (summa cum laude) in applied sciences from KU Leuven, Belgium, in 1996.

    He is currently a Full Professor with the Delft Center for Systems and Control, Delft University of Technology, The Netherlands. His research interests include learning and optimisation-based control of large-scale and hybrid systems, multilevel and distributed control, with applications in intelligent transportation and smart energy systems.

    Prof. De Schutter is a Senior Editor of the IEEE \textsc{Transactions on Intelligent Transportation Systems}.
\end{IEEEbiography}
\begin{IEEEbiography}[{\includegraphics[width=1in,height=1.25in,clip,keepaspectratio]{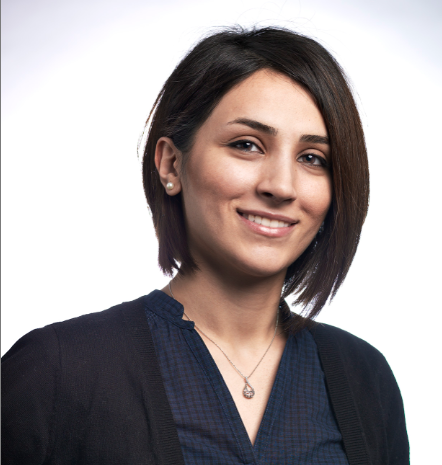}}]{Azita Dabiri}
    received the Ph.D. degree from the Automatic Control Group, Chalmers University of Technology, in 2016. She was a Post-Doctoral Researcher with the Department of Transport and Planning, Delft University of Technology, from 2017 to 2019. In 2019, she received an ERCIM Fellowship and also a Marie Curie Individual Fellowship, which allowed her to perform research at the Norwegian University of Technology (NTNU), as a Post-Doctoral Researcher, from 2019 to 2020, before joining the Delft Center for Systems and Control, TU Delft, as an Assistant Professor. Her research interests are in the areas of integration of model-based and learning-based control and its applications in transportation networks.
\end{IEEEbiography}
\vfill

\appendix  

We provide here additional insights on our proposed methodology and the numerical results. First, we provide a short investigation on the effects of different parametrisations of the MPC scheme \eqref{methodology:eq:mpc-func-approx:scheme}. Then, we report the evolution of all the parameters $\bm\theta$ during the learning process implemented in \Cref{sections:numerical}.
\begin{figure}[h]
    \centering
    \input{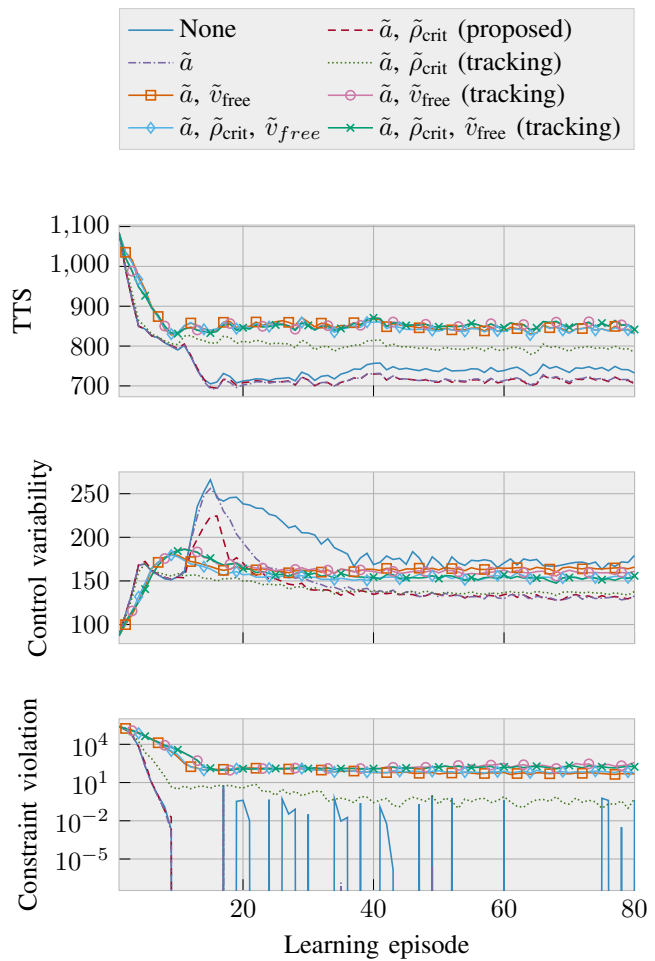}
    \caption{
        Evolution of the RL cost contributions for various different parametrisations $\bm\theta$. The legend lists for each entry the dynamics parameters that were allowed to be learnt in the corresponding simulation, including also the parametrisation that we proposed in this work, i.e., \textit{(proposed)}. If either $\tilde{\rho}_\text{crit}$, $\tilde{v}_\text{free}$ or both were also employed as tracking setpoints, the entry is marked accordingly, i.e., \textit{(tracking)}. To avoid visual clutter, only the average over the 15 simulation runs is shown for each parametrisation.
    }
    \label{appendix:fig:parametrisation-performances}
\end{figure}

\section*{Different MPC parametrisations}

Different parametrisations $\bm\theta$ of the MPC scheme \eqref{methodology:eq:mpc-func-approx:scheme} have been tested in simulations. In particular, we have considered the following variants:
\begin{itemize}
    \item learning a combination of the dynamics parameters $\tilde{a}$, $\tilde{v}_\text{free}$, and $\tilde{\rho}_\text{crit}$ (including the empty set, i.e., learning none of these parameters)
    \item employing the learnable dynamics parameters $\tilde{v}_\text{free}$ and $\tilde{\rho}_\text{crit}$ also as tracking setpoints of the stage cost \eqref{methodology:eq:mpc-func-approx:stage-cost} and terminal cost \eqref{methodology:eq:mpc-func-approx:terminal-cost}, i.e., $v_\text{sp} = \tilde{v}_\text{free}$, $\rho_\text{sp} = \tilde{\rho}_\text{crit}$.
\end{itemize}
The rationale behind trying out these various combinations is that, while having a richer parametrisation can potentially help in fitting more complex value functions, a drawback is that the learning task becomes much more complex and more likely to converge to a very suboptimal local minimum. Moreover, the sensitivity of the RL solution to the parametrisation can vary significantly from parameter to parameter, so learning some of them may end up in a less stable process than others. As is the case with most function approximators, the choice of parametrisation relies mostly on prior and/or expert knowledge (such as, in our case, the sensitivity analysis of the METANET model w.r.t. its parameters) and is mostly an iterative procedure (both for MPC and other function approximators), whose difficulty is attributable to the RL algorithm itself rather than to the proposed control scheme. However, meta-learning methods can be found in the literature to optimise over the selection of this and other hyper-parameters via, e.g., Bayesian Optimisation \cite{wu_2019_hyperparameter} and bilevel optimisation \cite{franceschi_2018_bilevel}.

\Cref{appendix:fig:parametrisation-performances} shows the influence on the performance of some of the combinations we tested. These tests were carried out with a lower variability of the randomly generated profiles in an effort to filter out the impact of randomness on the performance. One can notice that, when neither the dynamics parameters nor the tracking setpoints are learnt, the resulting controller converges to a solid performance. Adding $\tilde{a}$ to the set of learnable parameters seems to boost the performance even further, which can be attributed to the fact that, among the various parameters, $\tilde{a}$ is the one the METANET model is most sensitive to. Further testing indicates that learning also $\tilde{\rho}_\text{crit}$ achieves an additional performance improvement, while learning $\tilde{v}_\text{free}$ does not help. The last empirical finding is that making the tracking setpoints learnable, despite increasing the number of degrees of freedom of the parametrisation, does not bring any improvement.

\section*{Evolution of parametrisation}

Interested readers can find in \Cref{appendix:fig:learning-parameters-all} the evolution, during the whole learning process, of the whole parametrisation $\bm\theta$ of the MPC scheme \eqref{methodology:eq:mpc-func-approx:scheme}, as detailed in \Cref{sections:sub:mpc-func-approximator} and simulated in \Cref{sections:numerical}.
\begin{figure*}
    \centering
    \input{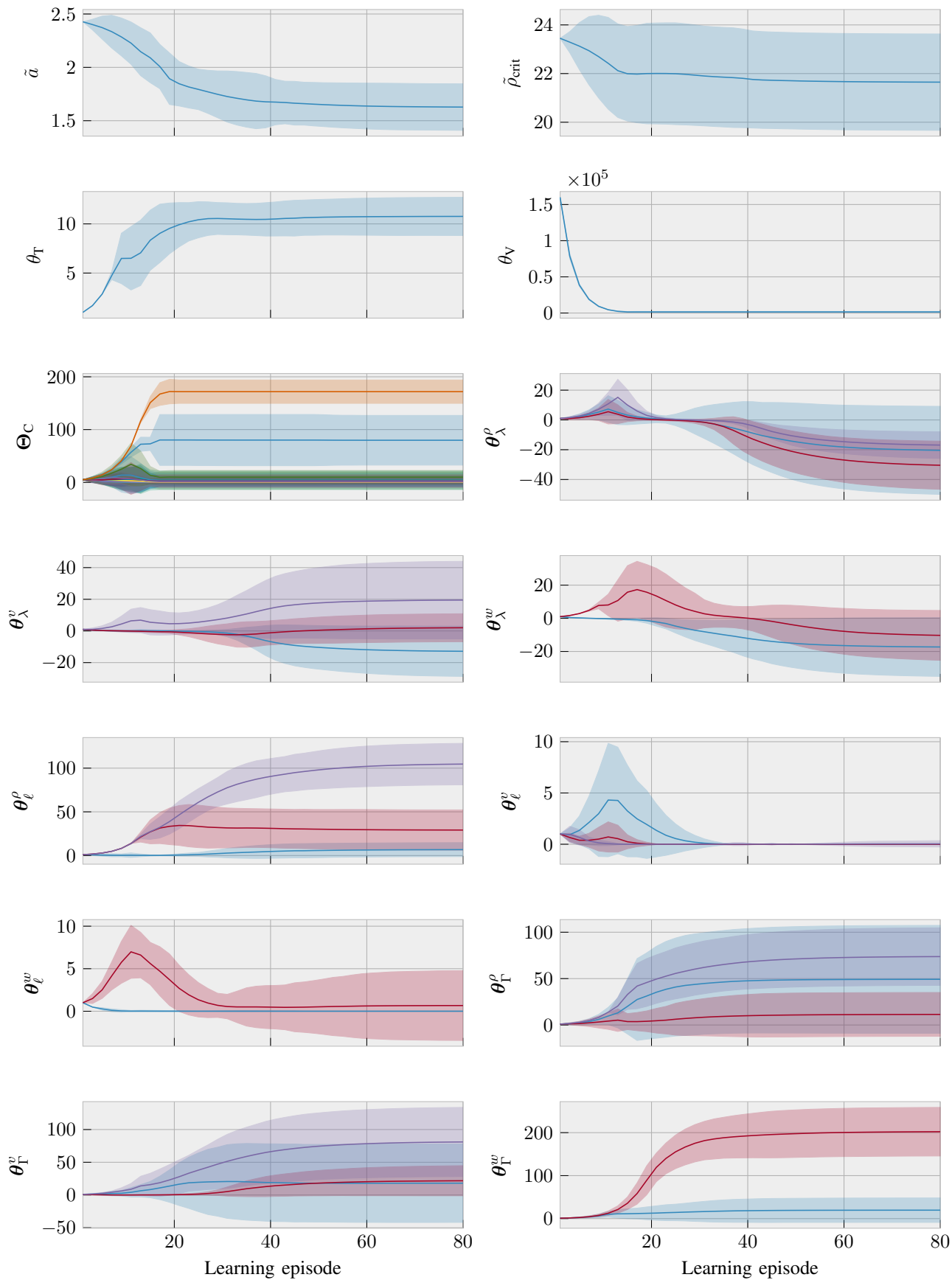}
    \caption{Average evolution of the parametrisation $\bm\theta$ during the learning process across the 15 simulation runs. For those parameters that are vector quantities, one colour is associated with each entry.}
    \label{appendix:fig:learning-parameters-all}
\end{figure*}

\end{document}